\documentclass[lettersize,journal]{IEEEtran}
\usepackage{amsmath,amsfonts,amssymb,amsthm}
\usepackage{xcolor}
\usepackage{algorithmic}
\usepackage{algorithm}
\usepackage{array}
\usepackage{subfig}
\usepackage{textcomp}
\usepackage{stfloats}
\usepackage{url}
\usepackage{verbatim}
\usepackage{graphicx}
\usepackage{cite}
\usepackage{multirow}
\def\BibTeX{{\rm B\kern-.05em{\sc i\kern-.025em b}\kern-.08em
    T\kern-.1667em\lower.7ex\hbox{E}\kern-.125emX}}
\usepackage{balance}

\def\cA{{\mathcal{A}}}  \def\cC{{\mathcal{C}}} 
\def\cE{{\mathcal{E}}}   \def\cH{{\mathcal{H}}}
   \def\cL{{\mathcal{L}}}
 \def\cN{{\mathcal{N}}} \def\cO{{\mathcal{O}}} \def\cP{{\mathcal{P}}}
 \def\cR{{\mathcal{R}}}

\def\diag{\mathop{\mathrm{diag}}}

\def\trace{\mathop{\mathrm{tr}}}

\def\bPhi{{\pmb{\Phi}}} \def\bphi{{\pmb{\phi}}}

\def\b0{{\pmb{0}}}\def\bLambda{{\pmb{\Lambda}}} 

\def\ba{{\mathbf{a}}} \def\bb{{\mathbf{b}}}  \def\bd{{\mathbf{d}}}
 \def\bff{{\mathbf{f}}} \def\bg{{\mathbf{g}}} \def\bh{{\mathbf{h}}}
   
\def\bm{{\mathbf{m}}} \def\bn{{\mathbf{n}}}  
 \def\br{{\mathbf{r}}}  
   \def\bx{{\mathbf{x}}}
\def\by{{\mathbf{y}}}   

\def\bA{{\mathbf{A}}} \def\bB{{\mathbf{B}}}  \def\bD{{\mathbf{D}}}
\def\bE{{\mathbf{E}}} \def\bF{{\mathbf{F}}} \def\bG{{\mathbf{G}}} \def\bH{{\mathbf{H}}}
\def\bI{{\mathbf{I}}}   
   \def\bP{{\mathbf{P}}}
 \def\bR{{\mathbf{R}}}  
\def\bU{{\mathbf{U}}} \def\bV{{\mathbf{V}}} \def\bW{{\mathbf{W}}}

\begin{document}
	
\title{WMMSE-Based Rate Maximization for RIS-Assisted MU-MIMO Systems}

\author{Hyuckjin Choi,~\IEEEmembership{Student Member,~IEEE,}\thanks{H. Choi and J. Choi are with the School of Electrical Engineering, Korea Advanced Institute of Science and Technology (e-mail: \{hugzin008,junil\}@kaist.ac.kr).} A. Lee Swindlehurst,~\IEEEmembership{Fellow,~IEEE,}\thanks{A. L. Swindlehurst is with the Center for Pervasive Communications and Computing, University of California, Irvine, CA 92697, USA (e-mail: swindle@uci.edu).} and Junil Choi,~\IEEEmembership{Senior Member,~IEEE.}
}



\maketitle

\begin{abstract}	
	Reconfigurable intelligent surface (RIS) technology, given its ability to favorably modify wireless communication environments, will play a pivotal role in the evolution of future communication systems. This paper proposes rate maximization techniques for both single-user and multiuser MIMO systems, based on the well-known weighted minimum mean square error (WMMSE) criterion. Using a suitable weight matrix, the WMMSE algorithm tackles an equivalent weighted mean square error (WMSE) minimization problem to achieve the sum-rate maximization. By considering a more practical RIS system model that employs a tensor-based representation enforced by the electromagnetic behavior exhibited by the RIS panel, we detail both the sum-rate maximizing and WMSE minimizing strategies for RIS phase shift optimization by deriving the closed-form gradient of the WMSE and the sum-rate with respect to the RIS phase shift vector. Our simulations reveal that the proposed rate maximization technique, rooted in the WMMSE algorithm, exhibits superior performance when compared to other benchmarks.
\end{abstract}


\section{Introduction}
\IEEEPARstart{T}{he} main objective of fifth and sixth generation (5G/6G) communication systems is to provide large signal bandwidths by targeting high-frequency carriers \cite{Andrews:2014,Wang:2014,Shafi:2017,Rappaport:2019,Tataria:2021}. However, high-frequency channels, with their strong line-of-sight (LoS) characteristics, face the issue of blockages \cite{Rappaport:2017}. This may necessitate the deployment of reconfigurable intelligent surfaces (RISs) that can offer cost-effective and energy-efficient solutions to enhance coverage \cite{Wu:2020,Huang:2020,Jian:2022}. An RIS, composed of a large number of passive elements, can adjust the direction of reflected signals and illuminate shadowed areas with low power consumption.

A multitude of RIS communication techniques, e.g. channel estimation, joint beamforming, and localization, are currently under extensive development \cite{Zheng:2019,Wang:2020,Wei:2021,Wu:2019,Zhang:2020,Cho:2021,Noh:2022,Kim:2022,Elzanaty:2021}. In particular, joint beamforming methods that simultaneously optimize the base station (BS) precoder and the RIS phase shifts have been devised \cite{Wu:2019,Zhang:2020,Cho:2021,Noh:2022,Kim:2022,Huang:2019,Shen:2019,Cui:2019,Pan:2020,You:2021,Abrardo:2021,Chen:2022}. Within the context of RIS-assisted communication systems, however, primary attention has been given to multiple-input single-output (MISO) systems in which the user terminals have only one antenna. For the MISO case, the receive (Rx) signal power can be converted into a quadratic form relative to the RIS phase shifts \cite{Wu:2019}, rendering problems more tractable. However, systems in which the users possess multiple Rx antennas present more significant challenges.

In rate maximization problems, the achievable rate for multiple-input multiple-output (MIMO) systems cannot be expressed in a quadratic form with respect to the RIS phase shift. Consequently, several heuristic techniques have been proposed, including steering the RIS with incidence and reflection angles \cite{Hong:2022}, the sum-path-gain-maximization (SPGM)-based algorithm \cite{Ning:2020}, and the iterative distance minimization (IDM) algorithm \cite{Zhou:2021}. While the RIS phase shifts are typically optimized iteratively, a stationary point can be evaluated element-wise for the rate maximization problem \cite{Zhang:2020,Kim:2022}. Machine learning (ML) approaches have also been exploited to optimize RIS systems \cite{Urakami:2023,Huang:2023,Taha:2020,Evmorfos:2023}.

The goal of the weighted minimum mean square error (WMMSE) algorithm is to design the transmit (Tx) beamformer to minimize the weighted mean squared error (WMSE), which is equivalent to maximizing the sum-rate for a particular choice of the weight matrix. In the joint Tx beamformer and RIS phase shift optimization, the majorization-minimization (MM) method was used to update the RIS phase shifts in \cite{Pan:2020,You:2021}. The stationary points of the WMSE were exploited for the RIS phase adjustment by ignoring the unit-modulus constraint in \cite{Abrardo:2021} or considering an upper bound on the WMMSE in \cite{Chen:2022}. These approaches however do not provide an optimal solution for the rate maximization problem since, unlike the Tx beamformer design, minimizing the WMSE does not necessarily maximize the sum-rate when it comes to the RIS phase shift optimization.

To achieve better sum-rate performance for multiuser MIMO (MU-MIMO) systems, in this paper we propose a sum-rate maximization technique. While using the WMMSE beamformer at the BS, the proposed RIS optimization method relies on the gradient of the sum-rate rather than the WMSE. Unlike using the WMSE stationary points in \cite{Abrardo:2021,Chen:2022}, setting the gradient of the sum-rate to zero in the proposed method always guarantees an increase in the sum-rate. Although finding the sum-rate stationary point poses a challenge, we tackle this problem by the gradient descent algorithm. It will be shown in Section~\ref{sec5} that the proposed method gives the best performance.

Another contribution of our work is that we develop our technique based on a generalized RIS system model that can be articulated through a tensor representation. Conventional RIS system models fall short in accommodating high-rank channels with respect to a single RIS element. In contrast, the tensor expression can deal with RIS channels whose rank is greater than one, a plausible scenario provided by an examination of RIS operational characteristics \cite{Najafi:2021}. For the tensor-based channel model, previous WMSE minimization techniques \cite{Pan:2020,Abrardo:2021} cannot be directly applied. Thus, in addition to the sum-rate maximization technique, we derive the best WMSE minimization technique for the tensor-based channel model. Moreover, as an example of an ML-based approach, we implement a convolutional neural network (CNN)-based algorithm for our scenario of interest. The contributions of this paper are summarized as follows:
\begin{enumerate}
	\item A generalized tensor-based RIS system model is proposed, and techniques for finding the optimal RIS configuration are developed within the generalized system model.
	\item The gradient of the sum-rate with respect to the RIS phase shift vector is derived for the tensor-based RIS system model.
    \item The WMSE minimization method is derived for the tensor-based RIS system model.
	\item Experimental results are provided to compare the two WMMSE-based algorithms, one that maximizes the sum rate and another that minimizes the WMSE, together with other benchmarks.
\end{enumerate}

The paper is structured as follows. Section~\ref{sec2} first develops the generalized tensor-based channel model for RIS-assisted MU-MIMO systems and defines the sum-rate maximization problem. Section~\ref{sec3} provides detailed derivations of the proposed techniques that maximize the sum-rate or minimize the WMSE based on the WMMSE criterion. We also describe the CNN-based approach in this section. Section~\ref{sec4} derives techniques for the special case of SU-MIMO systems, and Section~\ref{sec5} showcases various experimental results for the rate maximization techniques for both SU-MIMO and MU-MIMO systems. Concluding remarks follow in Section~\ref{sec6}.

\textbf{Notation:} Lower- and upper-case boldface letters are used to represent column vectors and matrices, respectively, while calligraphic upper-case boldface letters are tensors. The conjugate, inverse, transpose, and conjugate (Hermitian) transpose of matrix $\bA$ are $\bA^*$, $\bA^{-1}$, $\bA^\mathrm{T}$, and $\bA^\mathrm{H}$, respectively. The transpose of the matrix inverse $\bA^{-1}$ is $\bA^{-\mathrm{T}}$. The determinant and trace of matrix $\bA$ are denoted as $\det(\bA)$ and $\trace(\bA)$. The diagonal matrix whose diagonal elements are formed by the vector $\ba$ is denoted as $\diag\{\ba\}$. The $N\times N$ identity matrix is represented as $\bI_N$. The symbol $\otimes$ represents the Kronecker product. The notation $\text{Unif.}[x_a,x_b]$ indicates that a random variable $X$ is uniformly distributed in \mbox{$x_a\leq X \leq x_b$}. The complex multivariate Gaussian distribution for a vector whose elements are independent and have identical variance is written as $\cC\cN(\bm,\sigma^2\bI_N)$, with mean vector $\bm\in\mathbb{C}^{N\times 1}$ and variance $\sigma^2$.

Multiplication between a tensor $\boldsymbol{\cA}\in\mathbb{C}^{M\times N\times L}$ and a matrix $\bB\in\mathbb{C}^{N\times K}$ is defined as $\boldsymbol{\cA}\times_i \bB\in\mathbb{C}^{M\times K\times L}$ (in this case, $i=2$), where the $i$-th dimension of tensor $\boldsymbol{\cA}$ is combined with the first dimension of matrix $\bB$. The multi-dimensional Hadamard product $\odot_i$ combines the $i$-th dimension of $\bA\in\mathbb{C}^{M\times N}$ and the first dimension of $\bB\in\mathbb{C}^{N\times L}$ as $\bA\odot_i\bB\in\mathbb{C}^{M\times N\times L}$ (in this case, $i=2$), where the $n$-th slab of $\bA\odot_2\bB$ is the outer product of the $n$-th column of $\bA$ and the $n$-th row of $\bB$. The dimension shrinkage of tensor $\boldsymbol{\cA}\in\mathbb{C}^{M\times 1\times L}$ is denoted as $[\![\boldsymbol{\cA}]\!]\in\mathbb{C}^{M\times L}$, where dimensions with size one disappear.

\begin{figure}[!t]
	\centering
	\begin{center}$
		\begin{array}{c}
			\subfloat[Deployment of BS, RIS, and UEs.]{\includegraphics[width=\columnwidth]{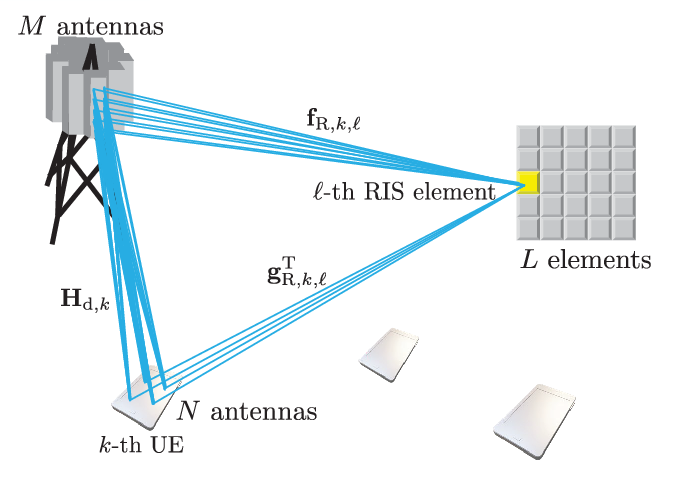} \label{subfig:1_a}}\\
			\subfloat[Angles of departure and arrival.]{\includegraphics[width=0.7\columnwidth]{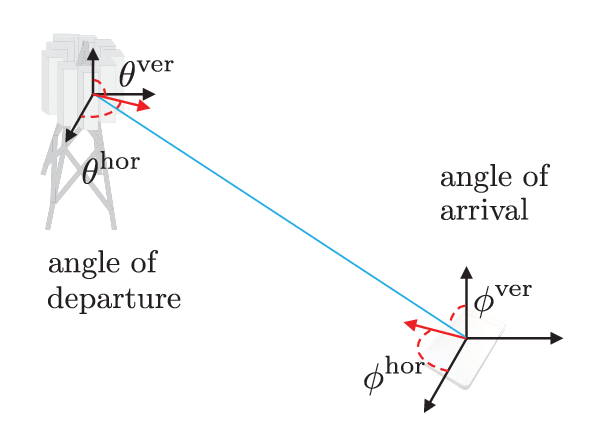} \label{subfig:1_b}}
		\end{array}$
	\end{center}
	\caption{System model considered in this paper. The BS has $M$ antennas, the RIS has $L$ elements, and $K$ UEs have $N$ antennas each.}
	\label{fig:system_model}
\end{figure}

\section{System Model and Problem Formulation}
\label{sec2}

We first provide the system and channel models for the direct and RIS-assisted channels in Section~\ref{sec2_1}. We briefly discuss the scope of this work in Section~\ref{sec2_2} and define the problem of interest in Section~\ref{sec2_3}.

\subsection{System and Channel Models}
\label{sec2_1}
We focus on a single-cell downlink system with a single RIS and $K$ users (UEs), as shown in Fig.~\ref{fig:system_model}~\subref{subfig:1_a}. The BS has $M$ antennas, each UE has $N$ antennas, and the RIS has $L$ purely passive elements. We consider a multi-path channel model for the direct and RIS channels. The direct channel $\bH_{\mathrm{d},k}\in\mathbb{C}^{M\times N}$ for the $k$-th UE is modeled as
\begin{align}\label{eq:dir_ch}
	\bH_{\mathrm{d},k}=\sum_{p_\mathrm{a}= 1}^{P_{\mathrm{a},k}}\gamma_{k,p_\mathrm{a}}\ba_M(\theta_{k,p_\mathrm{a}}^\text{hor},\theta_{k,p_\mathrm{a}}^\text{ver})\ba_N^\mathrm{H}(\phi_{k,p_\mathrm{a}}^\text{hor},\phi_{k,p_\mathrm{a}}^\text{ver}),
\end{align}
which consists of $P_{\mathrm{a},k}$ channel paths between the BS and the $k$-th UE. The path gain in \eqref{eq:dir_ch} is $\gamma_{k,p_\mathrm{a}}$. As shown in Fig.~\ref{fig:system_model}~\subref{subfig:1_b}, azimuth/elevation angles of arrival and azimuth/elevation angles of departure for the $p_\mathrm{a}$-th channel path are given as $\theta_{k,p_\mathrm{a}}^{\text{hor}}$, $\theta_{k,p_\mathrm{a}}^{\text{ver}}$, $\phi_{k,p_\mathrm{a}}^{\text{hor}}$, and $\phi_{k,p_\mathrm{a}}^{\text{ver}}$, respectively. We assume that the path gain follows the complex Gaussian distribution $\gamma_{k,p_\mathrm{a}}\sim\cC\cN(0,\beta_{\text{BU}}(d_{\text{BU},k,p_\mathrm{a}}/d_{\text{BU},\text{ref}})^{-\alpha_{\text{BU},p_\mathrm{a}}})$, where $d_{\text{BU},k,p_\mathrm{a}}$ is the distance of the $p_\mathrm{a}$-th channel path from the BS to the $k$-th UE, $\beta_{\text{BU}}$ is the path gain at the reference distance $d_{\text{BU},\text{ref}}$, and $\alpha_{\text{BU},p_\mathrm{a}}$ is the pathloss exponent for the $p_\mathrm{a}$-th channel path. The BS and UE antennas and the RIS elements are assumed to be configured with a uniform planar array (UPA) structure. For the BS, the array response vector is defined as
\begin{align}
	&\ba_M(\theta^\text{hor},\theta^\text{ver}) \nonumber \\
	&\ =[1\ e^{j\frac{2\pi d}{\lambda}\cos(\theta^\text{hor})\sin(\theta^\text{ver})}\ \cdots\ e^{j\frac{2\pi d}{\lambda}({M_\text{hor}}-1)\cos(\theta^\text{hor})\sin(\theta^\text{ver})}]^\mathrm{T} \nonumber \\
	&\ \quad\otimes[1\ e^{j\frac{2\pi d}{\lambda}\cos(\theta^\text{ver})}\ \cdots\ e^{j\frac{2\pi d}{\lambda}({M_\text{ver}}-1)\cos(\theta^\text{ver})}]^\mathrm{T},
\end{align}
where $\lambda$ represents the signal wavelength, and $d$ denotes the antenna spacing, which is assumed to be $d=\frac{\lambda}{2}$ in this paper. The number of antennas in the horizontal and vertical directions are $M_\text{hor}$ and $M_\text{ver}$, respectively, and the total number of BS antennas is $M=M_\text{hor}\times M_\text{ver}$. The UPA vectors $\ba_N(\theta^\text{hor},\theta^\text{ver})$ and $\ba_L(\theta^\text{hor},\theta^\text{ver})$ for the UE and RIS are similarly defined with $N=N_\text{hor}\times N_\text{ver}$ and $L=L_\text{hor}\times L_\text{ver}$.

Most previous works have relied on the following RIS-assisted channel model for the $k$-th UE:
\begin{align}\label{eq:r1}
	\bH_{\mathrm{R},k}^\text{eff}=\bF_{\mathrm{R}}\bPhi\bG_{\mathrm{R},k},
\end{align}
where $\bF_{\mathrm{R}}\in\mathbb{C}^{M\times L}$ defines the BS-to-RIS channel, $\bG_{\mathrm{R},k}\in\mathbb{C}^{L\times N}$ is the RIS-to-UE channel, and $\bPhi=\diag\{\bphi\}=\diag{[\phi_1\ \cdots\ \phi_L]^\mathrm{T}}$ represents the RIS phase shift matrix. Since the BS-to-RIS channel is shared by all UEs, the BS-to-RIS channel is not written with the UE \mbox{index $k$}. We can also express the conventional model in \eqref{eq:r1} as
\begin{align}\label{eq:r2}
	\bH_{\mathrm{R},k}^\text{eff}=\sum_{\ell=1}^L\phi_\ell\bff_{\mathrm{R},\ell}\bg_{\mathrm{R},k,\ell}^\mathrm{T},
\end{align}
where the $\ell$-th column of $\bF_{\mathrm{R}}$ is $\bff_{\mathrm{R},\ell}$, and the $\ell$-th row of $\bG_{\mathrm{R},k}$ is $\bg_{\mathrm{R},k,\ell}^\mathrm{T}$. 

Since the BS-to-RIS channel $\bff_{\mathrm{R},\ell}$ and the RIS-to-UE channel $\bg_{\mathrm{R},k,\ell}$ for the $k$-th UE and the $\ell$-th RIS element are the sum of channel paths, the RIS-assisted channel can be further expressed as
\begin{align}
    \bH_{\mathrm{R},k}^\text{eff}&=
    \sum_{\ell=1}^L\phi_\ell\left(\sum_{p_b=1}^{P_b}\bff_{\mathrm{R},\ell,p_b}\right)
    \left(\sum_{p_c=1}^{P_{c,k}}\bg_{\mathrm{R},k,\ell,p_c}^\mathrm{T}\right) \nonumber \\
    &= \sum_{\ell=1}^L\phi_\ell\left(\sum_{p_b=1}^{P_b}\sum_{p_c=1}^{P_{c,k}}
    \bff_{\mathrm{R},\ell,p_b}\bg_{\mathrm{R},k,\ell,p_c}^\mathrm{T}\right),
\end{align}
where $P_\mathrm{b}$ is the number of channel paths between the BS and RIS, and $P_{\mathrm{c},k}$ is the number of channel paths between the RIS and $k$-th UE. The BS-to-RIS channel $\bff_{\mathrm{R},\ell}$ and the RIS-to-UE channel $\bg_{\mathrm{R},k,\ell}$ consist of $P_b$ and $P_{c,k}$ channel paths, respectively. However, the conventional channel model has not taken into account the RIS physical response $\Omega_{p_b,p_c}$, which can vary for each pair of incoming and outgoing channel paths \cite{Najafi:2021}. With the RIS physical response $\Omega_{p_b,p_c}$, the RIS-assisted channel is given as
\begin{align}
    \bH_{\mathrm{R},k}^\text{eff}
    = \sum_{\ell=1}^L\phi_\ell\left(\sum_{p_b=1}^{P_b}\sum_{p_c=1}^{P_{c,k}}
    \Omega_{p_b,p_c}
    \bff_{\mathrm{R},\ell,p_b}\bg_{\mathrm{R},k,\ell,p_c}^\mathrm{T}\right),
\end{align}
where the summations over $p_b$ and $p_c$ are no longer separable. By denoting $\bH_{\mathrm{R},k,\ell}=\sum_{p_b=1}^{P_b}\sum_{p_c=1}^{P_{c,k}}\Omega_{p_b,p_c}\bff_{\mathrm{R},\ell,p_b}\bg_{\mathrm{R},k,\ell,p_c}^\mathrm{T}$, the RIS-assisted channel in \eqref{eq:r2} can be rewritten as
\begin{align}\label{eq:r3}
	\bH_{\mathrm{R},k}^\text{eff}=\sum_{\ell=1}^L\phi_\ell\bH_{\mathrm{R},k,\ell},
\end{align}
where the $\ell$-th RIS channel $\bH_{\mathrm{R},k,\ell}$ in \eqref{eq:r3} is no longer a rank-one matrix for $P_b>1$ and $P_{c,k}>1$, which means that the conventional model in \eqref{eq:r1} is not generally applicable. Only when the RIS physical response $\Omega_{p_b,p_c}$ is constant for different pairs of $(p_b,p_c)$ will the rank of $\bH_{\mathrm{R},k,\ell}$ equal one.

To provide a more generalized representation, we define an RIS channel tensor $\boldsymbol{\cH}_{\mathrm{R},k}$, whose $\ell$-th slab $\bH_{\mathrm{R},k,\ell}$ is given by $[\![\boldsymbol{\cH}_{\mathrm{R},k}(:,\ell,:)]\!]=\bH_{\mathrm{R},k,\ell}$. Using the RIS channel tensor, we can represent the RIS-assisted channel in \eqref{eq:r3} as
\begin{align}\label{eq:r4}
	\bH_{\mathrm{R},k}^\text{eff}=[\![\boldsymbol{\cH}_{\mathrm{R},k}\times_2\bphi]\!].
\end{align}
In cases where the RIS channel $\bH_{\mathrm{R},k,\ell}$ associated with the $\ell$-th RIS element has rank greater than one, the tensor-based representation in \eqref{eq:r4} offers a tractable algebraic formulation.

Without considering the RIS physical response $\Omega_{p_b,p_c}$, the RIS channel tensor $\boldsymbol{\cH}_{\mathrm{R},k}$ for the $k$-th UE can be expressed as
\begin{align}\label{eq:ris_ch_conv}
	\boldsymbol{\cH}_{\mathrm{R},k} &= \left(\sum_{p_\mathrm{b}= 1}^{P_\mathrm{b}} \gamma_{p_\mathrm{b}}  \ba_M(\theta_{p_\mathrm{b}}^\text{hor},\theta_{p_\mathrm{b}}^\text{ver})\ba_L^\mathrm{H}(\phi_{p_\mathrm{b}}^\text{hor},\phi_{p_\mathrm{b}}^\text{ver})\right)\nonumber \\
	&\qquad\odot_2 \left(\sum_{p_\mathrm{c}= 1}^{P_{\mathrm{c},k}}\gamma_{k,p_\mathrm{c}} \ba_L(\theta_{k,p_\mathrm{c}}^\text{hor},\theta_{k,p_\mathrm{c}}^\text{ver})\ba_N^\mathrm{H}(\phi_{k,p_\mathrm{c}}^\text{hor},\phi_{k,p_\mathrm{c}}^\text{ver})\right),
\end{align}
where the path gains $\gamma_{p_\mathrm{b}}$ and $\gamma_{k,p_\mathrm{c}}$ are distributed as $\gamma_{p_\mathrm{b}}\sim\cC\cN(0,\beta_{\text{BR}}(d_{\text{BR},p_\mathrm{b}}/d_{\text{BR},\text{ref}})^{-\alpha_{\text{BR},p_\mathrm{b}}})$ and $\gamma_{k,p_\mathrm{c}}\sim\cC\cN(0,\beta_{\text{RU}}(d_{\text{RU},k,p_\mathrm{c}}/d_{\text{RU},\text{ref}})^{-\alpha_{\text{RU},p_\mathrm{c}}})$. The meanings of these variables are consistent with the variables explained after \eqref{eq:dir_ch}.

If \eqref{eq:ris_ch_conv} holds, the $\ell$-th slab of RIS channel $[\![\boldsymbol{\cH}_{\mathrm{R},k}(:,\ell,:)]\!]$ $=\bH_{\mathrm{R},k,\ell}$ becomes a rank-one matrix, which boils down to the conventional model in \eqref{eq:r1}. As mentioned above, however, the RIS physical response should be modeled as $\Omega(\phi_{p_\mathrm{b}}^\text{hor},\phi_{p_\mathrm{b}}^\text{ver},\theta_{k,p_\mathrm{c}}^\text{hor},\theta_{k,p_\mathrm{c}}^\text{ver})$ as defined\footnote{Since the exact form of $\Omega(\phi_{p_\mathrm{b}}^\text{hor},\phi_{p_\mathrm{b}}^\text{ver},\theta_{k,p_\mathrm{c}}^\text{hor},\theta_{k,p_\mathrm{c}}^\text{ver})$ is complicated, we omit it and refer to \cite{Najafi:2021}.} in equations (10) and (11) of \cite{Najafi:2021}, since the RIS reacts differently to signals with different angles of arrival $\phi_{p_\mathrm{b}}^\text{hor}$ and $\phi_{p_\mathrm{b}}^\text{ver}$ and angles of departure $\theta_{k,p_\mathrm{c}}^\text{hor}$ and $\theta_{k,p_\mathrm{c}}^\text{ver}$. Given this more general RIS response, the channel tensor $\boldsymbol{\cH}_{\mathrm{R},k}$ is formed as
\begin{align}\label{eq:ris_ch_tens}
	\boldsymbol{\cH}_{\mathrm{R},k} =& \sum_{p_\mathrm{b}=1}^{P_\mathrm{b}}\sum_{p_\mathrm{c}=1}^{P_{\mathrm{c},k}}\gamma_{k,p_\mathrm{b},p_\mathrm{c}}
    \Omega(\phi_{p_\mathrm{b}}^\text{hor},\phi_{p_\mathrm{b}}^\text{ver},\theta_{k,p_\mathrm{c}}^\text{hor},\theta_{k,p_\mathrm{c}}^\text{ver}) \nonumber \\
    &\qquad\times\left(\ba_M(\theta_{p_\mathrm{b}}^\text{hor},\theta_{p_\mathrm{b}}^\text{ver})\ba_L^\mathrm{H}(\phi_{p_\mathrm{b}}^\text{hor},\phi_{p_\mathrm{b}}^\text{ver})\right) \nonumber \\
    &\qquad\qquad\odot_2\left(\ba_L(\theta_{k,p_\mathrm{c}}^\text{hor},\theta_{k,p_\mathrm{c}}^\text{ver})\ba_N^\mathrm{H}(\phi_{k,p_\mathrm{c}}^\text{hor},\phi_{k,p_\mathrm{c}}^\text{ver})\right),
\end{align}
where the gain of each channel path $\gamma_{k,p_\mathrm{b},p_\mathrm{c}}$ is distributed as $\gamma_{k,p_\mathrm{b},p_\mathrm{c}}\sim\cC\cN(0,\beta_{\text{BR}}(d_{\text{BR},p_\mathrm{b}}/d_{\text{BR},\text{ref}})^{-\alpha_{\text{BR},p_\mathrm{b}}}+\beta_{\text{RU}}(d_{\text{RU},k,p_\mathrm{c}}/d_{\text{RU},\text{ref}})^{-\alpha_{\text{RU},p_\mathrm{c}}})$. Note that the rank of the $\ell$-th slab of the RIS channel in \eqref{eq:ris_ch_tens} cannot be greater than $P_\mathrm{b}P_{\mathrm{c},k}$.


With the RIS channel tensor in \eqref{eq:r4}, the overall channel to the $k$-th UE can be formulated as
\begin{align}\label{eq:RISchannel}
	\bH_k=\bH_{\mathrm{d},k}+\bH_{\mathrm{R},k}^\text{eff}=\bH_{\mathrm{d},k}+[\![\boldsymbol{\cH}_{\mathrm{R},k}\times_2\bphi]\!].
\end{align}
The Rx signal at the $k$-th UE can thus be expressed as
\begin{align}\label{eq:system_model}
	\by_k=\bH_k^\mathrm{H}\bx+\bn_k=(\bH_{\mathrm{d},k}+[\![\boldsymbol{\cH}_{\mathrm{R},k}\times_2\bphi]\!])^\mathrm{H}\bx+\bn_k,
\end{align}
where the Tx signal at the BS is denoted by $\bx$, and the noise at the $k$-th UE is represented by $\bn_k \sim \cC\cN(\b0,\sigma_n^2\bI_N)$ with variance $\sigma_n^2$. The Tx signal $\bx$ consists of the data symbols for all UEs, which can be written as
\begin{align}
	\bx = \sum_{k=1}^K\bx_k = \sum_{k=1}^K\bB_k\bd_k=\bB\bd,
\end{align}
where $\bB_k\in\mathbb{C}^{M\times N}$ and  $\bd_k\in\mathbb{C}^{N\times 1}$ are the beamformer and the data symbols for the $k$-th UE. The overall beamformer matrix and symbol vector are given as $\bB=[\bB_1\ \cdots\ \bB_K]$ and $\bd=[\bd_1^\mathrm{T}\ \cdots\ \bd_K^\mathrm{T}]^\mathrm{T}$. The data symbols are assumed to be uncorrelated: $\mathbb{E}[\bd\bd^\mathrm{H}]=\bI_{KN}$.

\subsection{MISO/SIMO Systems}
\label{sec2_2}
When the UE has a single antenna, the conventional RIS-assisted channel in \eqref{eq:r1} can be represented as
\begin{align}\label{eq:MISOmodel}
	\bh_{\mathrm{R},k}^\text{eff}=\bF_{\mathrm{R},k}\bPhi\bg_{\mathrm{R},k}^\mathrm{T}=\bF_{\mathrm{R},k}\diag\{\bg_{\mathrm{R},k}\}\bphi.
\end{align}
Denoting the RIS channel without the RIS phase shift as $\bH_{\mathrm{R},k}=\bF_{\mathrm{R},k}\diag\{\bg_{\mathrm{R},k}\}$, the RIS-assisted channel in \eqref{eq:MISOmodel} becomes
\begin{align}
	\bh_{\mathrm{R},k}^\text{eff}=\bH_{\mathrm{R},k}\bphi,
\end{align}
which is the same as \eqref{eq:r4} since the third dimension of $\boldsymbol{\cH}_{\mathrm{R},k}$ is unity. This clearly shows that, even when the RIS reacts selectively to each channel path, we can still rely on the conventional RIS channel model in \eqref{eq:r1} for MISO systems. This also holds for single-input multiple-output (SIMO) systems. Thus, we develop RIS optimization techniques only for the MIMO case taking the tensor-based RIS channel model in \eqref{eq:r4} and \eqref{eq:ris_ch_tens} into account.

\subsection{Problem Objective}
\label{sec2_3}
For the system model in \eqref{eq:system_model}, our interest is to maximize the sum-rate $\cR=\sum_{k=1}^K\cR_k$ by solving the following problem
\begin{align}
	\text{(P1)}: \max_{\bB_1,\cdots,\bB_K,\bphi} &\ \cR \nonumber \\
	\text{s.t.}\ \quad &\ \text{Tr}(\bB\bB^\mathrm{H})=E_\text{tx}, \nonumber \\
	&\ |\phi_\ell|^2=1,\ \forall\ell\in\{1,\dots,L\}, \nonumber
\end{align}
where $E_\text{tx}$ is the total Tx power at the BS, and the RIS phase shift element $\phi_\ell$ has unit modulus. The achievable rate for the $k$-th UE is defined as \cite{heath2018foundations}
\begin{align}\label{eq:rate_k}
	\cR_k&=\log\det(\bI_N+\bR_{\tilde{n}_k}^{-1}\bH_k^\mathrm{H}\bR_{\bx_k}\bH_k) \nonumber \\
	&=\log\det(\bI_N+\bB_k^\mathrm{H}\bH_k\bR_{\tilde{n}_k}^{-1}\bH_k^\mathrm{H}\bB_k),
\end{align}
where the covariance matrix of the Tx signal for the $k$-th UE is $\bR_{\bx_k}=\mathbb{E}[\bx_k\bx_k^\mathrm{H}]=\bB_k\bB_k^\mathrm{H}$, and the covariance matrix of the inter-user interference and noise terms is evaluated as
\begin{align}\label{eq:efftv_no}
	\bR_{\tilde{n}_k}=\sum_{i\neq k}\bH_k^\mathrm{H}\bB_i\bB_i^\mathrm{H}\bH_k+\sigma_n^2\bI_N.
\end{align}
The $k$-th UE channel $\bH_k$ in \eqref{eq:rate_k} and \eqref{eq:efftv_no} is given in \eqref{eq:RISchannel}.

In the proposed method, we tackle (P1) by alternately optimizing the beamformers $\bB_1,\dots,\bB_K$ and the RIS phase shift vector $\bphi$. The WMMSE algorithm is a popular approach for solving the sum-rate maximization problem \cite{Christensen:2008,Shi:2011}. In the following sections, we develop RIS optimization methods exploiting the WMMSE algorithm.

\section{WMMSE-Based RIS phase shift optimization}
\label{sec3}
To maximize the sum-rate, the RIS phase shift vector $\bphi$ should be optimized jointly with the beamformer $\bB$. We exploit the well-known WMMSE beamformer for $\bB$. Specifically, at each iteration of the WMMSE algorithm, the beamformer for the $k$-th UE, before the power normalization, is computed as \cite{Christensen:2008}
\begin{align}\label{eq:WMMSEbf1}
	\bar{\bB}_k=&\left(\sum_{i=1}^K\bH_i\bA_i^\mathrm{H}\bW_i\bA_i\bH_i^\mathrm{H}+\frac{\sum_{i=1}^K\text{Tr}(\bA_i^\mathrm{H}\bW_i\bA_i)}{E_\text{tx}/\sigma_n^2}\bI_M\right)^{-1} \nonumber \\
	&\quad\times\bH_k\bA_k^\mathrm{H}\bW_k,
\end{align}
which is the solution of the WMMSE problem. In \eqref{eq:WMMSEbf1}, the MMSE filter $\bA_k\in\mathbb{C}^{N\times N}$ and the weight matrix $\bW_k\in\mathbb{C}^{N\times N}$ are given as
\begin{align}\label{eq:WMMSEa}
	&\bA_k=\bB_k^\mathrm{H}\bH_k\left(\sum_{i=1}^K\bH_k^\mathrm{H}\bB_i\bB_i^\mathrm{H}\bH_k+\sigma_n^2\bI_N\right)^{-1}
\end{align}
and
\begin{align} \label{eq:WMMSEw}
	&\bW_k=\bI_N+\bB_k^\mathrm{H}\bH_k\bR_{\tilde{n}_k}^{-1}\bH_k^\mathrm{H}\bB_k.
\end{align}
Note that each iteration of the WMMSE beamformer necessitates computation of the MMSE filter $\bA_k$ and weight matrix $\bW_k$ for all $k$. Once the beamformers for all UEs are determined, the power normalization factor is given by $b=\sqrt{\frac{E_\text{tx}}{\text{Tr}(\bar{\bB}\bar{\bB}^\mathrm{H})}}$, where $\bar{\bB}=[\bar{\bB}_1\ \cdots\ \bar{\bB}_K]$, and the WMMSE beamformer becomes
\begin{align}\label{eq:WMMSEbf2}
	\bB^\text{WMMSE}_k=b\bar{\bB}_k.
\end{align}

While the main objective of the WMMSE beamformer is to maximize the sum-rate, the RIS phase shift vector $\bphi$ can be designed to either maximize the sum-rate or minimize the WMSE. Section~\ref{sec3_2} develops a GD-based technique for maximizing the sum-rate, and Section~\ref{sec3_1} tackles the problem of minimizing the WMSE. Section~\ref{sec3_3} introduces a CNN-based approach to represent the performance of ML-based techniques which are being widely considered for future wireless communication systems. 

\subsection{Sum-Rate Maximization}
\label{sec3_2}
To maximize the sum-rate with the GD algorithm, we need to derive the gradient of the sum-rate with respect to $\bphi$. We first derive a recursive form of the sum-rate gradient that will be used for the algorithm's iterative updates. Then, we obtain the initial gradient, which then fully defines the recursive form.

\subsubsection{Gradient of the Achievable Rate}
\label{sec3_2_1}
To evaluate the gradient, we reformulate the achievable rate for the $k$-th UE in \eqref{eq:rate_k} as
\begin{align}\label{eq:rate_k1}
	\cR_k&=\log\det\Big(\bI_N+\bB_k^\mathrm{H}(\bH_{\mathrm{d},k}+[\![\boldsymbol{\cH}_{\mathrm{R},k}\times_2\bphi]\!])\bR_{\tilde{n}_k}^{-1} \nonumber \\
	&\qquad\qquad\quad \times (\bH_{\mathrm{d},k}+[\![\boldsymbol{\cH}_{\mathrm{R},k}\times_2\bphi]\!])^\mathrm{H}\bB_k\Big) \nonumber \\
	&=\log\det\Big(\bI_N+\bB_k^\mathrm{H}[\![\boldsymbol{\cH}_k\times_2\boldsymbol{\psi}]\!]\bR_{\tilde{n}_k}^{-1} \nonumber \\
	&\qquad\qquad\quad \times [\![\boldsymbol{\cH}_k\times_2\boldsymbol{\psi}]\!]^\mathrm{H}\bB_k\Big) \nonumber \\
	&=\log\det\Big(\bI_N+[\![\tilde{\boldsymbol{\cH}}_k\times_2\boldsymbol{\psi}]\!]\bR_{\tilde{n}_k}^{-1} [\![\tilde{\boldsymbol{\cH}}_k\times_2\boldsymbol{\psi}]\!]^\mathrm{H}\Big) \nonumber \\
\end{align}
where the concatenated channel tensor is given as 
\begin{align}
	\boldsymbol{\cH}_k=[\bH_{\mathrm{d},k}:\boldsymbol{\cH}_{\mathrm{R},k}]\in\mathbb{C}^{M\times(L+1)\times N},
\end{align}
and the concatenated RIS phase shift vector is written as
\begin{align}\label{eq:catRIS}
	\boldsymbol{\psi}=\begin{bmatrix}1 & \bphi^\mathrm{T}\end{bmatrix}^\mathrm{T}\in\mathbb{C}^{(L+1)\times1}.
\end{align}
These quantities enable us to write [\![$\boldsymbol{\cH}_k\times_2\boldsymbol{\psi}]\!]=\bH_{\mathrm{d},k}+[\![\boldsymbol{\cH}_{\mathrm{R},k}\times_2\bphi]\!]$. To simplify the following derivations, we define the effective channel as
\begin{align} 
	\tilde{\boldsymbol{\cH}}_k=\boldsymbol{\cH}_k\times_1\bB_k^\ast\in\mathbb{C}^{N\times(L+1)\times N}.
\end{align}
The following property is useful to further describe the achievable rate for the $k$-th UE.
    
\noindent\textit{\textbf{Property 1:}} The product of the three-dimensional tensor $\boldsymbol{\cA}$ and vector $\bb$ in the second dimension of $\boldsymbol{\cA}$ can be represented as
\begin{align}
    [\![\boldsymbol{\cA}\times_2\bb]\!]&=\begin{bmatrix}[\![\boldsymbol{\cA}(:,:,1)]\!]\bb & [\![\boldsymbol{\cA}(:,:,N)]\!]\bb & \cdots \end{bmatrix} \nonumber \\
    &=\begin{bmatrix}[\![\boldsymbol{\cA}(:,:,1)]\!] & [\![\boldsymbol{\cA}(:,:,N)]\!] & \cdots \end{bmatrix}(\bI\otimes\bb), \nonumber
\end{align}
and the following also holds:
\begin{align}
    [\![\boldsymbol{\cA}\times_2\bb]\!]&=\begin{bmatrix}[\![\boldsymbol{\cA}(1,:,:)]\!]^\mathrm{T}\bb & [\![\boldsymbol{\cA}(2,:,:)]\!]^\mathrm{T}\bb & \cdots\end{bmatrix}^\mathrm{T} \nonumber \\
    &=(\bI\otimes\bb^\mathrm{T})\begin{bmatrix}[\![\boldsymbol{\cA}(1,:,:)]\!]^\mathrm{T} & [\![\boldsymbol{\cA}(2,:,:)]\!]^\mathrm{T} & \cdots\end{bmatrix}^\mathrm{T}, \nonumber
\end{align}
indicating that the block-expanded vector $(\bI\otimes\bb)$ can be shifted to either the left or right side.

Using Property 1 with $\bar{\bH}_{k,n}=$ \mbox{$[\![\tilde{\boldsymbol{\cH}}_k(n,:,:)]\!]$}. The rate for the $k$-th UE can be represented as
\begin{align}\label{eq:rate_k3}
    \cR_k&=\log\det\Big\{\bI_N+\begin{bmatrix}\bar{\bH}_{k,1}^\mathrm{T}\boldsymbol{\psi} & \cdots & \bar{\bH}_{k,N}^\mathrm{T}\boldsymbol{\psi}\end{bmatrix}^\mathrm{T} \nonumber \\
	&\qquad\qquad\ \ \times\bR_{\tilde{n}_k}^{-1}\begin{bmatrix}\bar{\bH}_{k,1}^\mathrm{H}\boldsymbol{\psi}^\ast & \cdots & \bar{\bH}_{k,N}^\mathrm{H}\boldsymbol{\psi}^\ast\end{bmatrix}\Big\}.
\end{align}
Before expanding \eqref{eq:rate_k3} further, we first state the following lemma.

\noindent{\textit{\textbf{Lemma 1:}} For any arbitrary matrices $\bA$ and $\bB$ with proper dimensions, the following holds:}
	\begin{align}
		\det\left(\bI+\bA^\mathrm{T}\bB\bA\right)=\prod_{m=1}^M\left(1+\ba_m^\mathrm{T}\bP_m^\bot\ba_m\right), \nonumber
	\end{align}
	where $\bA=[\ba_1\ \cdots\ \ba_M]$, and 
	\begin{align}
		\bP_m^\bot=\begin{cases}
			\bB &,\ m=1,\\
			\bP_{m-1}^\bot-\frac{\bP_{m-1}^\bot\ba_{m-1}\ba_{m-1}^\mathrm{T}\bP_{m-1}^\bot}{1+\ba_{m-1}^\mathrm{T}\bP_{m-1}^\bot\ba_{m-1}} &,\ m\neq 1.
		\end{cases} \nonumber
	\end{align}
	\begin{proof}The proof is provided in Appendix A.\end{proof}

Using Lemma 1, $\cR_k$ in \eqref{eq:rate_k3} can be transformed into a semi-quadratic form as
\begin{align}\label{eq:rate_k2}
	\cR_k=\sum_{n=1}^N\log\left(1+\boldsymbol{\psi}^\mathrm{T}\bar{\bH}_{k,n}\bP_{k,n}^\bot\bar{\bH}_{k,n}^\mathrm{H}\boldsymbol{\psi}^\ast\right),
\end{align}
where the matrix $\bP_{k,n}^\bot$ is defined recursively as
\begin{align} \label{eq:mat_P}
	\bP_{k,n}^\bot = \begin{cases}\bR_{\tilde{n}_k}^{-1} &,\ n=1,\\
		\bP_{k,n-1}^\bot-\frac{\bP_{k,n-1}^\bot\bar{\bH}_{k,n-1}^\mathrm{H}\boldsymbol{\psi}^\ast\boldsymbol{\psi}^\mathrm{T}\bar{\bH}_{k,n-1}\bP_{k,n-1}^\bot}{1+\boldsymbol{\psi}^\mathrm{T}\bar{\bH}_{k,n-1}\bP_{k,n-1}^\bot\bar{\bH}_{k,n-1}^\mathrm{H}\boldsymbol{\psi}^\ast}&,\ n\neq 1.
	\end{cases}
\end{align}
Introducing an auxiliary function $q_{k,n}^{i,j}(\boldsymbol{\psi})=\boldsymbol{\psi}^\mathrm{T}\bar{\bH}_{k,i}\bP_{k,n}^\bot\bar{\bH}_{k,j}^\mathrm{H}\boldsymbol{\psi}^\ast$, the achievable rate in \eqref{eq:rate_k2} can be again rewritten as
\begin{align}
	\cR_k=\sum_{n=1}^N\log\left(1+q_{k,n}^{n,n}(\boldsymbol{\psi})\right).
\end{align}
Due to its recursive structure, the auxiliary function $q_{k,n}^{i,j}(\boldsymbol{\psi})$ can be represented as
\begin{align}\label{eq:func_q}
	&q_{k,n}^{i,j}(\boldsymbol{\psi}) \nonumber \\
	&\ =\boldsymbol{\psi}^\mathrm{T}\bar{\bH}_{k,i}\bP_{k,n}^\bot\bar{\bH}_{k,j}^\mathrm{H}\boldsymbol{\psi}^\ast \nonumber \\
	&\ =\boldsymbol{\psi}^\mathrm{T}\bar{\bH}_{k,i}\bP_{k,n-1}^\bot\bar{\bH}_{k,j}^\mathrm{H}\boldsymbol{\psi}^\ast \nonumber \\
	&\ \ -\frac{\boldsymbol{\psi}^\mathrm{T}\bar{\bH}_{k,i}\bP_{k,n-1}^\bot\bar{\bH}_{k,n-1}^\mathrm{H}\boldsymbol{\psi}^\ast\boldsymbol{\psi}^\mathrm{T}\bar{\bH}_{k,n-1}\bP_{k,n-1}^\bot\bar{\bH}_{k,j}^\mathrm{H}\boldsymbol{\psi}^\ast}{1+\boldsymbol{\psi}^\mathrm{T}\bar{\bH}_{k,n-1}\bP_{k,n-1}^\bot\bar{\bH}_{k,n-1}^\mathrm{H}\boldsymbol{\psi}^\ast} \nonumber \\
	&\ =q_{k,n-1}^{i,j}(\boldsymbol{\psi})-\frac{q_{k,n-1}^{i,n-1}(\boldsymbol{\psi})q_{k,n-1}^{n-1,j}(\boldsymbol{\psi})}{1+q_{k,n-1}^{n-1,n-1}(\boldsymbol{\psi})}.
\end{align}
From \eqref{eq:rate_k2}, the gradient of $\cR_k$ can be evaluated as
\begin{align}
	\nabla_{\boldsymbol{\psi}}\cR_k=\sum_{n=1}^N\nabla_{\boldsymbol{\psi}}\log\left(1+q_{k,n}^{n,n}(\boldsymbol{\psi})\right)
	=\sum_{n=1}^N\frac{\nabla_{\boldsymbol{\psi}}q_{k,n}^{n,n}(\boldsymbol{\psi})}{1+q_{k,n}^{n,n}(\boldsymbol{\psi})},
\end{align}
where $\nabla_{\boldsymbol{\psi}}q_{k,n}^{i,j}(\boldsymbol{\psi})$ is derived using \eqref{eq:func_q}:
\begin{align}\label{eq:grad_q_r}
	&\nabla_{\boldsymbol{\psi}}q_{k,n}^{i,j}(\boldsymbol{\psi}) \nonumber \\
	&\quad=\nabla_{\boldsymbol{\psi}}q_{k,n-1}^{i,j}(\boldsymbol{\psi})-\frac{q_{k,n-1}^{i,n-1}(\boldsymbol{\psi})}{1+q_{k,n-1}^{n-1,n-1}(\boldsymbol{\psi})}\nabla_{\boldsymbol{\psi}}q_{k,n-1}^{n-1,j}(\boldsymbol{\psi}) \nonumber \\
	&\quad\quad-\frac{q_{k,n-1}^{n-1,j}(\boldsymbol{\psi}_)}{1+q_{k,n-1}^{n-1,n-1}(\boldsymbol{\psi})}\nabla_{\boldsymbol{\psi}}q_{k,n-1}^{i,n-1}(\boldsymbol{\psi}) \nonumber \\
	&\quad\quad+\frac{q_{k,n-1}^{i,n-1}(\boldsymbol{\psi})q_{k,n-1}^{n-1,j}(\boldsymbol{\psi})}{\left(1+q_{k,n-1}^{n-1,n-1}(\boldsymbol{\psi})\right)^2}\nabla_{\boldsymbol{\psi}}q_{k,n-1}^{n-1,n-1}(\boldsymbol{\psi}).
\end{align}
The gradient of $q_{k,n}^{i,j}(\boldsymbol{\psi})$ in \eqref{eq:grad_q_r} clearly shows that we need to evaluate $\nabla_{\boldsymbol{\psi}}q_{k,n-1}^{i,j}$, $\nabla_{\boldsymbol{\psi}}q_{k,n-1}^{n-1,j}$, $\nabla_{\boldsymbol{\psi}}q_{k,n-1}^{i,n-1}$, and $\nabla_{\boldsymbol{\psi}}q_{k,n-1}^{n-1,n-1}$ to compute $\nabla_{\boldsymbol{\psi}}q_{k,n}^{n,n}$. The first gradient $\nabla_{\boldsymbol{\psi}}q_{k,1}^{i,j}$ is derived in the following section.

\subsubsection{Evaluation of First Gradient}
\label{sec3_2_2}
The first auxiliary function $q_{k,1}^{i,j}(\boldsymbol{\psi})=\boldsymbol{\psi}^\mathrm{T}\bar{\bH}_{i,1}\bR_{\tilde{n}_k}^{-1}\bar{\bH}_{j,1}^\mathrm{T}\boldsymbol{\psi}^\ast$ is a fractional function of $\boldsymbol{\psi}$, where $\bR_{\tilde{n}_k}^{-1}$ is also a function of $\boldsymbol{\psi}$. The gradient of $q_{k,1}^{i,j}(\boldsymbol{\psi})$ is then evaluated as
\begin{align}\label{eq:grad_q}
	&\nabla_{\boldsymbol{\psi}}q_{k,1}^{i,j}(\boldsymbol{\psi}) \nonumber \\
	&=\nabla_{\boldsymbol{\psi}}\left(\boldsymbol{\psi}^\mathrm{T}\bar{\bH}_{k,i}\bR_{\tilde{n}_k}^{-1}\bar{\bH}_{k,j}^\mathrm{H}\boldsymbol{\psi}^\ast\right) \nonumber \\
	&=\bar{\bH}_{k,i}\bR_{\tilde{n}_k}^{-1}\bar{\bH}_{k,j}^\mathrm{H}\boldsymbol{\psi}^\ast+[\![\nabla_{\boldsymbol{\psi}}\bR_{\tilde{n}_k}^{-1}\times_1 \bar{\bH}_{k,i}^\mathrm{T}\boldsymbol{\psi}]\!]\bar{\bH}_{k,j}^\mathrm{H}\boldsymbol{\psi}^\ast \nonumber \\
	&\stackrel{(a)}{=}\bar{\bH}_{k,i}\bR_{\tilde{n}_k}^{-1}\bar{\bH}_{k,j}^\mathrm{H}\boldsymbol{\psi}^\ast \nonumber \\
	&\quad -[\![\nabla_{\boldsymbol{\psi}}\bR_{\tilde{n}_k}\times_1 \bR_{\tilde{n}_k}^{-\mathrm{T}}\bar{\bH}_{k,i}^\mathrm{T}\boldsymbol{\psi}]\!]\bR_{\tilde{n}_k}^{-1}\bar{\bH}_{k,j}^\mathrm{H}\boldsymbol{\psi}^\ast \nonumber \\
	&\stackrel{(b)}{=}\bar{\bH}_{k,i}\bR_{\tilde{n}_k}^{-1}\bar{\bH}_{k,j}^\mathrm{H}\boldsymbol{\psi}^\ast \nonumber \\
	&\quad -[\![\left[\frac{\partial}{\partial \psi_1}\bR_{\tilde{n}_k}\ :\ \cdots\ :\ \frac{\partial}{\partial \psi_{L+1}}\bR_{\tilde{n}_k}\right]\times_1 \bR_{\tilde{n}_k}^{-\mathrm{T}}\bar{\bH}_{k,i}^\mathrm{T}\boldsymbol{\psi}]\!] \nonumber \\
	&\quad\quad\ \ \times\bR_{\tilde{n}_k}^{-1}\bar{\bH}_{k,j}^\mathrm{H}\boldsymbol{\psi}^\ast ,
\end{align}
where (a) holds because $\nabla_{\boldsymbol{\psi}}\bR_{\tilde{n}_k}^{-1}=-\left(\nabla_{\boldsymbol{\psi}}\bR_{\tilde{n}_k}\times_1 \bR_{\tilde{n}_k}^{-T}\right)\times_3\bR_{\tilde{n}_k}^{-1}$, and (b) is from
\begin{align}
	\nabla_{\boldsymbol{\psi}}\bR_{\tilde{n}_k}=\left[\frac{\partial}{\partial \psi_1}\bR_{\tilde{n}_k}\ :\ \cdots\ :\ \frac{\partial}{\partial \psi_{L+1}}\bR_{\tilde{n}_k}\right].
\end{align}
The partial derivative matrix $\frac{\partial}{\partial \psi_\ell}\bR_{\tilde{n}_k}\in\mathbb{C}^{N\times N}$ is the $\ell$-th slab of $\nabla_{\boldsymbol{\psi}}\bR_{\tilde{n}_k}\in\mathbb{C}^{N\times (L+1)\times N}$.

To derive $\nabla_{\boldsymbol{\psi}}\bR_{\tilde{n}_k}$, we reformulate the effective noise covariance matrix $\bR_{\tilde{n}_k}$ as
\begin{align}\label{eq:efftv_no1}
	\bR_{\tilde{n}_k}&=\sigma_n^2\bI_N+\sum_{i\neq k}(\bH_{\mathrm{d},k}+[\![\boldsymbol{\cH}_{\mathrm{R},k}\times_2\bphi]\!])^\mathrm{H}\bB_i \nonumber \\
	&\qquad\qquad\qquad\ \ \times\bB_i^\mathrm{H}(\bH_{\mathrm{d},k}+[\![\boldsymbol{\cH}_{\mathrm{R},k}\times_2\bphi]\!]) \nonumber \\
	&= \sigma_n^2\bI_N+\sum_{i\neq k}[\![\boldsymbol{\cH}_k\times_2\boldsymbol{\psi}]\!]^\mathrm{H}\bB_i\bB_i^\mathrm{H}[\![\boldsymbol{\cH}_k\times_2\boldsymbol{\psi}]\!] \nonumber \\
	&= \sigma_n^2\bI_N+\sum_{i\neq k}[\![\tilde{\boldsymbol{\cH}}_{k,i}\times_2\boldsymbol{\psi}]\!]^\mathrm{H}[\![\tilde{\boldsymbol{\cH}}_{k,i}\times_2\boldsymbol{\psi}]\!] \nonumber \\
	&= \sigma_n^2\bI_N+\sum_{i\neq k}\begin{bmatrix}\tilde{\bH}_{k,i,1}\boldsymbol{\psi} & \cdots \tilde{\bH}_{k,i,N}\boldsymbol{\psi}
	\end{bmatrix}^\mathrm{H}\nonumber \\
	&\qquad\qquad\qquad\quad\times\begin{bmatrix}\tilde{\bH}_{k,i,1}\boldsymbol{\psi} & \cdots \tilde{\bH}_{k,i,N}\boldsymbol{\psi}
	\end{bmatrix} \nonumber \\
	&= \sigma_n^2\bI_N+(\bI_N\otimes\boldsymbol{\psi}^\mathrm{H})\sum_{i\neq k}\begin{bmatrix}\tilde{\bH}_{k,i,1} & \cdots & \tilde{\bH}_{k,i,N}\end{bmatrix}^\mathrm{H} \nonumber \\
	&\qquad\quad\ \ \times\begin{bmatrix}\tilde{\bH}_{k,i,1} & \cdots & \tilde{\bH}_{k,i,N}\end{bmatrix}(\bI_N\otimes\boldsymbol{\psi}).
\end{align}
Most of the derivation in \eqref{eq:efftv_no1} follows the procedure in \eqref{eq:rate_k1}. The dimension-reduced matrix $\tilde{\bH}_{k,i,n}$ in \eqref{eq:efftv_no1} is defined as $\tilde{\bH}_{k,i,n}=[\![\tilde{\boldsymbol{\cH}}_{k,i}(:,:,n)]\!]$. The partial derivative of the effective noise covariance matrix in \eqref{eq:efftv_no1} is given as
\begin{align}\label{eq:grad_R}
	\frac{\partial}{\partial \psi_\ell}\bR_{\tilde{n}_k}=&(\bI_N\otimes\boldsymbol{\psi}^\mathrm{H})\sum_{i\neq k}\begin{bmatrix}\tilde{\bH}_{k,i,1} & \cdots & \tilde{\bH}_{k,i,N}\end{bmatrix}^\mathrm{H} \nonumber \\
	&\qquad\ \ \times\begin{bmatrix}\tilde{\bH}_{k,i,1} & \cdots & \tilde{\bH}_{k,i,N}\end{bmatrix}(\bI_N\otimes\mathbf{e}_\ell),
\end{align} 
where $\mathbf{e}_\ell$ is the unit vector with one only at the $\ell$-th position and zeros elsewhere.

As we evaluate the gradient of the first auxiliary function $q_{k,1}^{i,j}(\boldsymbol{\psi})$, all the terms $\nabla_{\boldsymbol{\psi}}q_{k,n}^{n,n}$ ($n\in\{1,\dots,N\}$) necessary to compute the gradient of the achievable rate $\nabla_{\boldsymbol{\psi}}\cR_k$ for the $k$-th UE can be generated, which completes the derivation of gradient of the sum-rate $\nabla_{\boldsymbol{\psi}}\cR=\sum_{k=1}^K\nabla_{\boldsymbol{\psi}}\cR_k$.

With the gradient of the sum-rate and the learning rate $\beta$, the concatenated RIS phase shift vector $\boldsymbol{\psi}$ can be updated as
\begin{align}\label{eq:GDupdate}
	\boldsymbol{\psi}=\cP\left(\boldsymbol{\psi}+\beta(\nabla_{\boldsymbol{\psi}}\cR)^\ast\right),
\end{align} 
with the projection function $\cP(\boldsymbol{\psi})=\exp\{j\angle(\boldsymbol{\psi})\}$ such that $\boldsymbol{\psi}$ satisfies the unit-modulus constraint. Given the definition of $\boldsymbol{\psi}$ in \eqref{eq:catRIS}, the RIS phase shift vector $\bphi$ can be derived as
\begin{align}\label{eq:recover_phi}
	\bphi=\boldsymbol{\psi}_{2:(L+1)}/\psi_1,
\end{align}
where $\boldsymbol{\psi}_{2:(L+1)}$ denotes the vector containing the second to the \mbox{$(L+1)$}-th elements of $\boldsymbol{\psi}$. The sum-rate in \eqref{eq:rate_k2} preserves the optimality even when dividing $\boldsymbol{\psi}$ by $\psi_1$. We refer to the proposed GD method as MaxR-WMMSE, whose specifics are outlined in Algorithm~\ref{alg2}.

MaxR-WMMSE is the joint optimization algorithm for the BS beamformer $\bB$ and the RIS phase shifts $\bphi$. Since the sum-rate is a non-convex function, it is difficult to find the global optimum. Nevertheless, the convergence to a local optimum can be proved from the following
\begin{align}\label{eq:convg1}
    \cR(\bB^{(t+1)},\bphi^{(t)}) \geq \cR(\bB^{(t)},\bphi^{(t)})
\end{align}
and
\begin{align}\label{eq:convg2}
    \cR(\bB^{(t+1)},\bphi^{(t+1)}) \geq \cR(\bB^{(t+1)},\bphi^{(t)}).
\end{align}
Given RIS phase shifts $\bphi^{(t)}$, \eqref{eq:convg1} implies that the WMMSE algorithm leads to an increase in the rate in each iteration. Even under the unit-modulus constraint, MaxR-WMMSE converges to a local optimum since MaxR-WMMSE ensures \eqref{eq:convg2} as shown in the following theorem.

\noindent{\textit{\textbf{Theorem 1:}}} There always exists $\beta$ for the projection $\cP(\cdot)$ and the gradient $(\nabla_{\bphi}\cR)^\ast$ such that the following holds:
\begin{align}
    \cR(\cP(\bphi+\beta(\nabla_{\bphi}\cR)^\ast)) \geq \cR(\bphi).
\end{align}
\begin{proof} The proof is provided in Appendix B.
\end{proof}

By setting $\bphi^{(t+1)}=\cP(\bphi^{(t)}+\beta(\nabla_{{\bphi^{(t)}}}\cR)^\ast)$ with proper $\beta$, it is clear that \eqref{eq:convg2} holds for MaxR-WMMSE. 

\begin{algorithm}[t]
	\caption{MaxR-WMMSE.}\label{alg:alg1}
	\begin{algorithmic}
		\STATE \textsc{Initialize} $\bB_k,\ \forall k\in\{1,\dots,K\}$ and $\bphi$
		\STATE \textsc{repeat} 
		\STATE \hspace{0.4cm}$ \textbf{WMMSE beamformer:}\  \bB_k,\ \forall k\in\{1,\dots,K\}  $
\STATE \hspace{0.8cm}$ \bA_k \gets \eqref{eq:WMMSEa},\ \forall k\in\{1,\dots,K\} $
\STATE \hspace{0.7cm}$ \bW_k \gets \eqref{eq:WMMSEw},\ \forall k\in\{1,\dots,K\} $
\STATE \hspace{0.83cm}$ \bB_k \gets \eqref{eq:WMMSEbf1},\ \forall k\in\{1,\dots,K\} $
\STATE \hspace{0.4cm}$ \text{Normalize } \bB_1,\dots,\bB_K \text{ as in }\eqref{eq:WMMSEbf2}$
		\STATE
		\STATE \hspace{0.4cm} \textbf{RIS Gradient Update:}  $\bphi$
		\STATE \hspace{0.8cm} $ \nabla_{\boldsymbol{\psi}}\cR=0 $
		\STATE \hspace{0.8cm} \textbf{for } $ k=1:K $
		\STATE \hspace{1.3cm}$ q_{k,1}^{i,j} \gets$ \eqref{eq:grad_q}-\eqref{eq:grad_R}, $\forall i,j\in\{1,\dots,N\} $
		\STATE \hspace{1.3cm} \textbf{for } $ n=2:N $
		\STATE \hspace{1.8cm} $ \nabla_{\boldsymbol{\psi}}q_{k,n}^{i,j}(\boldsymbol{\psi})\gets \eqref{eq:grad_q_r},\ \forall i,j\in\{n,\dots,N\} $
		\STATE \hspace{1.3cm} \textbf{end}
		\STATE \hspace{1.3cm} $ \nabla_{\boldsymbol{\psi}}\cR\gets\nabla_{\boldsymbol{\psi}}\cR+\sum_{n=1}^N\frac{\nabla_{\boldsymbol{\psi}}q_{k,n}^{n,n}(\boldsymbol{\psi})}{1+q_{k,n}^{n,n}(\boldsymbol{\psi})} $
		\STATE \hspace{0.8cm} \textbf{end}
		\STATE
		\STATE \hspace{0.4cm} \textbf{Learning Rate Adaptation (bisection method)}
		\STATE \hspace{0.8cm} $\beta_\text{max}=100$,\ $\beta_\text{min}=0$
		\STATE \hspace{0.8cm} \textbf{for } $i=1:30$
		\STATE \hspace{1.3cm} $\beta=(\beta_\text{max}+\beta_\text{min})/2$
		\STATE \hspace{1.3cm} $\boldsymbol{\psi}_\text{new}=\exp\{j\angle(\boldsymbol{\psi}+\beta\nabla_{\boldsymbol{\psi}}\cR)\}$
		\STATE \hspace{1.3cm} \textbf{if } $\cR(\boldsymbol{\psi}_\text{new}) > \cR(\boldsymbol{\psi})$
		\STATE \hspace{1.8cm} $\beta_\text{min}=\beta$
		\STATE \hspace{1.3cm} \textbf{else}
		\STATE \hspace{1.8cm} $\beta_\text{max}=\beta$
		\STATE \hspace{1.3cm} \textbf{end}
		\STATE \hspace{0.8cm} \textbf{end}
		\STATE \hspace{0.8cm} $\boldsymbol{\psi}\gets\cP\left(\boldsymbol{\psi}+\beta(\nabla_{\boldsymbol{\psi}}\cR)^\ast\right)$
		\STATE \hspace{0.8cm} $\bphi=\boldsymbol{\psi}_{2:(L+1)}/\psi_1$
		\STATE {\textsc{until convergence}}
		\STATE {\textsc{return }} $\bB_k,\ \forall k\in\{1,\dots,K\}$ and $\bphi$
	\end{algorithmic}
	\label{alg2}
\end{algorithm}

\subsection{WMSE Minimization}
\label{sec3_1}
To define the WMSE, the symbol error matrix is first defined as
\begin{align}\label{eq:SymbErr}
	\bE_k&=(\hat{\bd}_k-\bd_k)(\hat{\bd}_k-\bd_k)^\mathrm{H} \nonumber \\
	&=(\bA_k\by_k-\bd_k)(\bA_k\by_k-\bd_k)^\mathrm{H},
\end{align}
where the symbol is estimated using the linear combiner $\bA_k$ as $\hat{\bd}_k=\bA_k\by_k$. Given~\eqref{eq:SymbErr}, the WMSE for the $k$-th UE is defined as
\begin{align}
	\cE_k=\text{Tr}\left\{\mathbb{E}[\bW_k\bE_k]\right\}.
\end{align}
Our goal is to choose the RIS phase shift vector $\bphi$ under the unit-modulus constraint in order to minimize the WMSE:
\begin{align}
	\text{(P2)}:\min_{\bphi} &\ \sum_{k=1}^K\cE_k \nonumber \\
	\text{s.t.}  &\ |\phi_\ell|^2=1,\ \forall\ell\in\{1,\dots,L\}. \nonumber
\end{align}
We solve (P2) using the Lagrangian dual formulation
\begin{align}
	\text{(P2')}:\min_{\bphi,\lambda_1,\dots,\lambda_L}\cL(\bphi,\lambda_1,\dots,\lambda_\ell), \nonumber
\end{align}
where the Lagrangian is defined as
\begin{align}
	\cL(\bphi,\lambda_1,\dots,\lambda_L)=\sum_{k=1}^K\cE_k+\sum_{\ell=1}^L\lambda_\ell(|\phi_\ell|^2-1).
\end{align}

The mean squared error (MSE) required to compute the objective in terms of the optimal RIS phase shift vector $\bphi$ can be written as
\begin{align}\label{eq:WMSE4}
	&\mathbb{E}[\bE_k]=\sum_{i=1}^K\begin{bmatrix}\bar{\bR}_{k,i,1}^\mathrm{H} & \bar{\bR}_{k,i,2}^\mathrm{H} & \cdots \end{bmatrix}(\bI\otimes\bphi^\ast) \nonumber \\
	&\qquad\qquad\quad\times(\bI\otimes\bphi^\mathrm{T})\begin{bmatrix}\bar{\bR}_{k,i,1}^\mathrm{T} & \bar{\bR}_{k,i,2}^\mathrm{T} & \cdots\end{bmatrix}^\mathrm{T} \nonumber \\
	&\qquad\quad+\sum_{i=1}^K(\bI\otimes\bphi^\mathrm{H})\begin{bmatrix}\tilde{\bR}_{k,i,1} & \tilde{\bR}_{k,i,2} & \cdots\end{bmatrix}^\mathrm{H}\tilde{\bH}_{k,i} \nonumber \\
	&\qquad\quad+\sum_{i=1}^K\tilde{\bH}_{k,i}^\mathrm{H}\begin{bmatrix}\tilde{\bR}_{k,i,1} & \tilde{\bR}_{k,i,2} & \cdots \end{bmatrix}(\bI\otimes\bphi) \nonumber \\
	&\qquad\quad-(\bI\otimes\bphi^\mathrm{H})\begin{bmatrix}\tilde{\bR}_{k,k,1} & \tilde{\bR}_{k,k,2} & \cdots\end{bmatrix}^\mathrm{H} \nonumber \\
	&\qquad\quad-\begin{bmatrix}\tilde{\bR}_{k,k,1} & \tilde{\bR}_{k,k,2} & \cdots\end{bmatrix}(\bI\otimes\bphi)+\cO_\bphi.
\end{align}
The detailed derivation of $\mathbb{E}[\bE_k]$ is given in Appendix~C.

The MSE for the $k$-th UE in \eqref{eq:WMSE4} is quadratic in the vector $\bphi$, which allows for the following compact WMSE expression:
\begin{align}\label{eq:WMSE5}
	&\cE_k=\sum_{n=1}^N\sum_{i=1}^K\bphi^\mathrm{T}\bar{\bR}_{k,i,n} \bW_k\bar{\bR}_{k,i,n}^\mathrm{H}\bphi^\ast \nonumber \\
	&\quad+\sum_{n=1}^N\sum_{i=1}^K\bphi^\mathrm{H} \tilde{\bR}_{k,i,n}^\mathrm{H}\tilde{\bH}_{k,i}\bW_k(:,n) \nonumber \\
	&\quad+\sum_{n=1}^N\sum_{i=1}^K\bW_k(n,:)\tilde{\bH}_{k,i}^\mathrm{H}\tilde{\bR}_{k,i,n}\bphi \nonumber \\
	&\quad-\sum_{n=1}^N\bphi^\mathrm{H}\tilde{\bR}_{k,k,n}^\mathrm{H}\bW_k(:,n)-\sum_{n=1}^N\bW_k(n,:)\tilde{\bR}_{k,k,n}\bphi +\cO_\bphi.
\end{align}
Given the WMSE in \eqref{eq:WMSE5}, the gradient of the Lagrangian $\cL(\bphi,\lambda_1,\dots,\lambda_L)$ in (P2') can be obtained as
\begin{align}\label{eq:LagrangeDerv}
	&\nabla_{\bphi^\ast}\cL(\bphi,\lambda_1,\dots,\lambda_L) \nonumber \\
	&=\nabla_{\bphi^\ast}\left\{\sum_{k=1}^K\cE_k+(\bPhi^\mathrm{H}\bPhi-\bI_L)\boldsymbol{\lambda}\right\} \nonumber \\
	&=\sum_{k=1}^K\left\{\sum_{n=1}^N\sum_{i=1}^K\bar{\bR}_{k,i,n}^\ast\bW_k^\mathrm{T}\bar{\bR}_{k,i,n}^\mathrm{T}\bphi\right. \nonumber \\
	&+\left.\sum_{n=1}^N\left(\sum_{i=1}^K\tilde{\bR}_{k,i,n}^\mathrm{H}\tilde{\bH}_{k,i}-\tilde{\bR}_{k,k,n}^\mathrm{H}\right)\bW_k(:,n)\right\}+\boldsymbol{\Lambda}\bphi,
\end{align}
where $\bLambda =\diag\{\boldsymbol{\lambda}\}$ with Lagrange multipliers $\boldsymbol{\lambda}=[\lambda_1\ \cdots\ \lambda_L]^\mathrm{T}$. From \eqref{eq:LagrangeDerv}, the unconstrained stationary point for (P2') is obtained as
\begin{align}\label{eq:WMSEupdate}
	&\bphi^\star= \left(\sum_{k=1}^K\sum_{n=1}^N\sum_{i=1}^K\bar{\bR}_{k,i,n}^\ast\bW_k^\mathrm{T}\bar{\bR}_{k,i,n}^\mathrm{T}+\bLambda\right)^{-1} \nonumber \\
	&\qquad\times\left\{\sum_{k=1}^K\sum_{n=1}^N\left(\tilde{\bR}_{k,k,n}^\mathrm{H}-\sum_{i=1}^K\tilde{\bR}_{k,i,n}^\mathrm{H}\tilde{\bH}_{k,i}\right)\bW_k(:,n)\right\}.
\end{align}

\begin{algorithm}[t]
	\caption{MinE-WMMSE.}\label{alg:alg1}
	\begin{algorithmic}
		\STATE \textsc{Initialize} $\bB_k,\ \forall k\in\{1,\dots,K\}$ and $\bphi$
		\STATE \textsc{repeat} 
		\STATE \hspace{0.4cm}$ \textbf{WMMSE beamformer:}\  \bB_k,\ \forall k\in\{1,\dots,K\}  $
		\STATE \hspace{0.8cm}$ \bA_k \gets \eqref{eq:WMMSEa},\ \forall k\in\{1,\dots,K\} $
		\STATE \hspace{0.7cm}$ \bW_k \gets \eqref{eq:WMMSEw},\ \forall k\in\{1,\dots,K\} $
		\STATE \hspace{0.83cm}$ \bB_k \gets \eqref{eq:WMMSEbf1},\ \forall k\in\{1,\dots,K\} $
		\STATE \hspace{0.4cm}$ \text{Normalize } \bB_1,\dots,\bB_K \text{ as in }\eqref{eq:WMMSEbf2}$
		\STATE
		\STATE \hspace{0.4cm} \textbf{RIS phase shift vector:}  $\bphi$
		\STATE \hspace{0.96cm} $ \bphi^\star \gets \eqref{eq:WMSEupdate},\eqref{eq:lambda} $
		\STATE \hspace{0.96cm} $\bphi = \cP(\bphi^\star)$
		\STATE {\textsc{until convergence}}
		\STATE {\textsc{return }} $\bB_k,\ \forall k\in\{1,\dots,K\}$ and $\bphi$
	\end{algorithmic}
	\label{alg1}
\end{algorithm}

\begin{figure*}[!t]
    \centering
    \includegraphics[width=1.55\columnwidth]{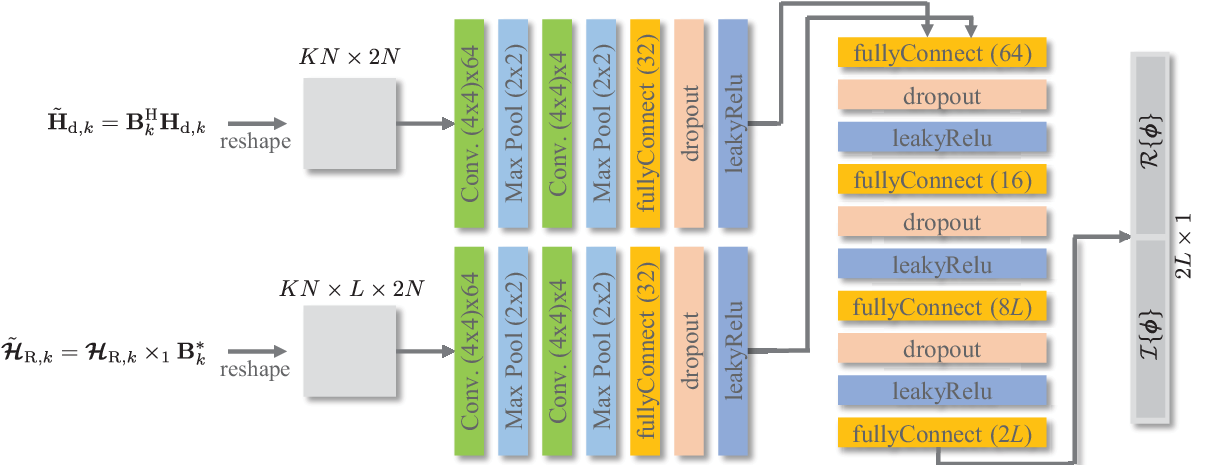}
    \caption{Structure of CNN-WMMSE network.}
    \label{fig:CNNstructure}
\end{figure*}

Since the elements of the optimized RIS phase shift vector $\bphi^\star$ will not in general adhere to the unit-modulus constraint, we modify the result as $\bphi=\cP(\bphi^\star)$. In the alternating optimization, the RIS phase shift vector $\bphi$ in \eqref{eq:WMSEupdate} is updated given estimates of the MMSE filter $\bA_k$, weight matrix $\bW_k$ of the WMMSE problem, and WMMSE beamformer $\bB_k$. Previous studies on WMSE minimization have varied in their methods for determining $\mathbf{\Lambda}$ \cite{Pan:2020,Abrardo:2021}. Our approach aligns with the method presented in \cite{Pan:2020}, which has been empirically established as the most effective strategy. The coefficient $\mathbf{\Lambda}$ is obtained as\cite{Pan:2020}
\begin{align}\label{eq:lambda}
    \bLambda = \frac{1}{\rho_\text{max}}\bI_L,
\end{align}
where $\rho_\text{max}$ is the largest eigenvalue of $\sum_{k=1}^K\sum_{n=1}^N$ $\sum_{i=1}^K\bar{\bR}_{k,i,n}^\ast\bW_k^\mathrm{T}\bar{\bR}_{k,i,n}^\mathrm{T}$. We refer to the alternating optimization as MinE-WMMSE, which is summarized in Algorithm~\ref{alg1}.

Regarding convergence, MinE-WMMSE does not always satisfy \eqref{eq:convg2} since minimizing the WMSE is not equivalent to maximizing the sum-rate for the given RIS phase shift. Instead, the convergence of MinE-WMMSE is numerically investigated in Section~\ref{sec5_2}.

\subsection{CNN-based Approach}
\label{sec3_3}
ML approaches have emerged as promising strategies for optimizing the RIS phase shifts, capitalizing on their ability to process extensive datasets and train complex algorithms. In this context, deep neural network (DNN) and reinforcement learning (RL) methodologies are predominantly used \cite{Urakami:2023,Huang:2023,Taha:2020,Evmorfos:2023}. The RL framework excels in environments where the action-policy is intricately tied to the evolving state, where the RIS phase shifts and channel state information (CSI) can be conceptualized as actions and states, respectively. Our interest in this paper, however, centers on rate maximization with perfect CSI. In this case, a DNN-based method proves to be more appropriate, and we employ a CNN-based approach, distinguished for its ability to distill channel features from the CSI.

\begin{figure}[!t]
    \centering
    \includegraphics[width=0.95\columnwidth]{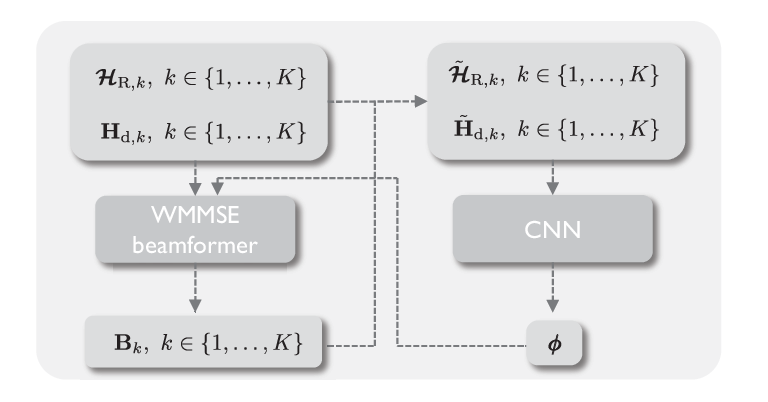}
    \caption{Flowchart of CNN-WMMSE algorithm.}
    \label{fig:CNNflowchart}
\end{figure}

The CNN structure we employed for the experiment is shown in Fig.~\ref{fig:CNNstructure}. The CNN structure takes the direct and RIS channels, after applying the beamformer $\bB_k$, as dual inputs. The fully-connected layers combine the features extracted from two convolutional layers and determine the RIS phase shifts. To optimize its performance, the CNN is jointly combined with the WMMSE beamformer as in Fig.~\ref{fig:CNNflowchart}. To maximize the sum-rate, the loss function is defined as
\begin{align}
    \cR(\bphi^{(t)})-\cR(\bphi^{(t+1)}),
\end{align}
where $\cR(\bphi^{(t)})$ and $\cR(\bphi^{(t+1)})$ are the previous and present sum-rates, respectively. We refer to the above CNN-based method as the CNN-WMMSE algorithm.

The CNN-WMMSE conducts the back propagation for every iteration of the WMMSE algorithm such that the network in Fig.~\ref{fig:CNNstructure} can adapt to the WMMSE beamformer update. While this methodology significantly enhances beamforming performance, it also introduces considerable overhead.

\section{Adaptation to SU-MIMO Systems}
\label{sec4}
Although the techniques described above were developed for the MU-MISO case, we show in this section that they can be easily adapted to the SU-MIMO scenario as well. In the numerical examples in Section~\ref{sec5_1}, we will show that the technique developed here works slightly better for the SU-MIMO case than previous state-of-the-art algorithms.
For SU-MIMO systems, (P1) can be represented as
\begin{align}
	\text{(P1')}: \max_{\bB,\bphi}\  &\ \log\det\left(\bI_N+\frac{1}{\sigma_n^2}\bB^\mathrm{H}\bH\bH^\mathrm{H}\bB\right) \nonumber \\
	\text{s.t.}\ \  &\ \text{Tr}(\bB\bB^\mathrm{H})=E_\text{tx}, \nonumber \\
	&\ |\phi_\ell|^2=1,\ \forall\ell\in\{1,\dots,L\}, \nonumber
\end{align}
where the beamformer $\bB$ and the channel $\bH$ are both $M\times N$ matrices. Since the channel $\bH$ is the sum of the direct channel $\bH_{\mathrm{d}}$ and the RIS channel $[\![\boldsymbol{\cH}_\mathrm{R}\times_2\bphi]\!]$, the channel $\bH$ changes with the RIS phase shift vector $\bphi$.

The singular value decomposition (SVD)-based beamformer with water-filling power allocation is known to be optimal for SU-MIMO systems \cite{tse2005fundamentals}. With the given RIS phase shift vector $\bphi$, the channel $\bH$ can be decomposed as $\bH = \bU\bD\bV^\mathrm{H}$, from which the optimal beamformer $\bB$ is given as
\begin{align}
	\bB^\star=\bU\sqrt{\text{diag}\{[p_1^\star\ \cdots\ p_N^\star]\}},
\end{align}
where the allocated power $[p_1^\star\ \cdots\ p_N^\star]$ satisfies the transmit power constraint in (P1') such that $\sum_{i=1}^N p_i=E_\text{tx}$. In many previous works, the RIS phase shift vector $\bphi$ is alternately optimized with the SVD-based beamformer \cite{Zhang:2020,Hong:2022,Ning:2020,Zhou:2021} to generate the system operating parameters.

For SU-MIMO systems, the WMMSE beamformer can still be evaluated using 
Eqs.~\eqref{eq:WMMSEbf1}-\eqref{eq:WMMSEbf2}. Thus, we provide two GD-based algorithms for comparison: GD-SVD based on the SVD beamformer, and GD-WMMSE based on the WMMSE beamformer. The RIS phase shifts are optimized with either beamformer, as detailed below.

For a given beamformer $\bB$, the achievable rate in (P1') can be written as
\begin{align}\label{eq:su_rate}
	\cR &=\log\det\bigg(\bI_N+\frac{1}{\sigma_n^2}\bB^\mathrm{H}(\bH_{\mathrm{d}}+[\![\boldsymbol{\cH}_\mathrm{R}\times_2\bphi]\!]) \nonumber \\&\qquad\qquad\quad\times(\bH_{\mathrm{d}}+[\![\boldsymbol{\cH}_\mathrm{R}\times_2\bphi]\!])^\mathrm{H}\bB\bigg) \nonumber \\
	&=\log\det\bigg(\bI_N+\frac{1}{\sigma_n^2}\bB^\mathrm{H}[\![\boldsymbol{\cH}\times_2\bphi]\!][\![\boldsymbol{\cH}\times_2\bphi]\!]^\mathrm{H}\bB\bigg) \nonumber \\
	&=\log\det\bigg(\bI_N+\frac{1}{\sigma_n^2}[\![\tilde{\boldsymbol{\cH}}\times_2\boldsymbol{\psi}]\!][\![\tilde{\boldsymbol{\cH}}\times_2\boldsymbol{\psi}]\!]^\mathrm{H}\bigg) \nonumber \\
	&=\sum_{n=1}^N\log\left(1+\boldsymbol{\psi}^\mathrm{T}\bar{\bH}_n\bP_n^\bot\bar{\bH}_n^\mathrm{H}\boldsymbol{\psi}^\ast\right),
\end{align}
where $\boldsymbol{\cH}=[\bH_{\mathrm{d}}:\boldsymbol{\cH}_\mathrm{R}]\in\mathbb{C}^{M\times(L+1)\times N}$ and \mbox{$\boldsymbol{\psi}=[1:\bphi]$} $\in\mathbb{C}^{(L+1)\times 1}$ are the concatenated tensor and vector, respectively. The channel tensor $\boldsymbol{\cH}$ and the beamformer matrix $\bB$ are combined as $\tilde{\boldsymbol{\cH}}=\boldsymbol{\cH}\times_1\bB^\ast$. The channel matrix $\bar{\bH}_n$ can be found from the channel tensor as $\bar{\bH}_n=[\![\tilde{\boldsymbol{\cH}}(n,:,:)]\!]$. The matrix $\bP_n^\bot$ in \eqref{eq:su_rate} is given as
\begin{align}
	\bP_{n}^\bot = \begin{cases}\frac{1}{\sigma_n^2}\bI_N &,\ n=1,\\
		\bP_{n-1}^\bot-\frac{\bP_{n-1}^\bot\bar{\bH}_{n-1}^\mathrm{H}\boldsymbol{\psi}^\ast\boldsymbol{\psi}^\mathrm{T}\bar{\bH}_{n-1}\bP_{n-1}^\bot}{1+\boldsymbol{\psi}^\mathrm{T}\bar{\bH}_{n-1}\bP_{n-1}^\bot\bar{\bH}_{n-1}^\mathrm{H}\boldsymbol{\psi}^\ast}&,\ n\neq 1.
	\end{cases}
\end{align}

As in Section~\ref{sec3_2}, the achievable rate $\cR$ can be succinctly expressed with the auxiliary function $q_n^{i,j}(\boldsymbol{\psi})=\boldsymbol{\psi}^\mathrm{T}\bar{\bH}_i\bP_n^\bot\bar{\bH}_j^\mathrm{H}\boldsymbol{\psi}^\ast$ as
\begin{align}
	\cR = \sum_{n=1}^N\log\left(1 + q_n^{n,n}(\boldsymbol{\psi})\right),
\end{align}
for which the gradient is evaluated as
\begin{align}\label{eq:su_grad_R}
	\nabla_{\boldsymbol{\psi}}\cR=\sum_{n=1}^N\frac{\nabla_{\boldsymbol{\psi}}q_n^{n,n}(\boldsymbol{\psi})}{1+q_n^{n,n}(\boldsymbol{\psi})}.
\end{align}
The auxiliary function $q_n^{i,j}(\boldsymbol{\psi})$ and its gradient $\nabla_{\boldsymbol{\psi}} q_n^{i,j}(\boldsymbol{\psi})$ have the same recursive structure as in \eqref{eq:func_q} and \eqref{eq:grad_q_r}, respectively. The first gradient of the auxiliary function $q_n^{i,j}(\boldsymbol{\psi})$ is evaluated as
\begin{align}
	\nabla_{\boldsymbol{\psi}}q_1^{i,j}(\boldsymbol{\psi})=\nabla_{\boldsymbol{\psi}}\frac{1}{\sigma_n^2}\boldsymbol{\psi}^\mathrm{T}\bar{\bH}_i\bar{\bH}_j^\mathrm{H}\boldsymbol{\psi}^\ast
	=\frac{1}{\sigma_n^2}\bar{\bH}_i\bar{\bH}_j^\mathrm{H}\boldsymbol{\psi}^\ast,
\end{align}
which is used to compute all auxiliary function values required for calculating the gradient in \eqref{eq:su_grad_R}. The gradient of the achievable rate updates the concatenated RIS phase shift vector $\boldsymbol{\psi}$ as
\begin{align}
	\boldsymbol{\psi}= \cP\left(\boldsymbol{\psi}+\beta(\nabla_{\boldsymbol{\psi}}\cR)^\ast\right),
\end{align}
and the RIS phase shift vector $\bphi$ can be recovered as $\bphi=\boldsymbol{\psi}_{2:(L+1)}/\psi_1$.

\begin{figure}[!t]
	\centering
	\includegraphics[width=0.9\columnwidth]{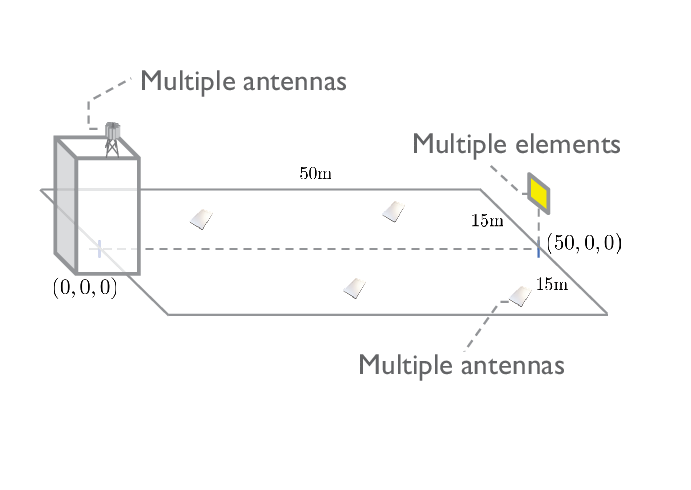}
	\caption{Simulation scenario for the SU-MIMO and MU-MIMO systems. The UEs are uniformly distributed in a \mbox{50m $\times$ 30m} area, where the height of BS is 35 m, and the height of RIS is 15 m.}
	\label{fig:simulation_scenario}
\end{figure}

\section{Simulation Results}
\label{sec5}

For the numerical studies, we consider the scenario shown in Fig.~\ref{fig:simulation_scenario}, where the BS is located at $(0,0,35)$ m, the RIS is at $(50,0,15)$ m, and the UEs are uniformly distributed within a 50m $\times$ 30m area. Given a UE location, the direct and RIS channels $\bH_\mathrm{d}$ and $\boldsymbol{\cH}_\mathrm{R}$ can be evaluated using \eqref{eq:dir_ch} and \eqref{eq:ris_ch_tens}. 
For the $k$-th UE, the indices associated with the LoS paths are $p_\mathrm{a}=1$ and $(p_\mathrm{b},p_\mathrm{c})=(1,1)$, and the indices for the NLoS paths are $\{2,\dots,P_{\mathrm{a},k}\}$ and $\{(p_\mathrm{b},p_\mathrm{c})|(p_\mathrm{b},p_\mathrm{c})\neq(1,1),p_\mathrm{b}\in\{1,\dots,P_\mathrm{b}\},p_\mathrm{c}\in\{1,\dots,P_{\mathrm{c},k}\}\}$. The distance of the LoS path $d_\text{LoS}$ is determined by the distance between the Tx and Rx nodes, and the distances associated with the NLoS paths are randomly generated as $d_\text{NLoS}=d_\text{LoS} + d_\mathrm{u}$, where $d_\mathrm{u}\sim\text{Unif.}[0,0.4d_\text{LoS}]$. The pathloss exponents are 2.5 for the LoS paths and 3.0 for the NLoS paths. The default number of channel paths is 16. The default Tx power is 30 dBm, and the noise variance $\sigma_n^2$ is -104 dBm for the simulations.

In the following sections, we present the performance of the proposed algorithms for both SU-MIMO and MU-MIMO scenarios. Section~\ref{sec5_1} applies various RIS optimization techniques specifically developed for the SU-MIMO case, and Section~\ref{sec5_2} provides analyses of the WMMSE-based algorithms for MU-MIMO systems.

\subsection{Experiments for SU-MIMO Systems}
\label{sec5_1}
For the SU-MIMO case, we assess the performance of several algorithms, including the element-wise (EW) update algorithms EW-TR \cite{Kim:2022} and EW-SV \cite{Zhang:2020}, the SPGM-based approach \cite{Ning:2020}, the WMSE-up method of \cite{Chen:2022}, IDM \cite{Zhou:2021}, and the array steering technique in \cite{Hong:2022}. We refer to our proposed GD methods based on the SVD and WMMSE beamformers as GD-SVD and GD-WMMSE, respectively. The EW algorithms derive a closed-form solution for the optimal RIS phase shift one RIS element at a time while holding the other RIS phase shifts fixed. Since EW-SV assumes the rank of the RIS channel $\bH_{\mathrm{R},k,\ell}$ in \eqref{eq:r3} is one, the solution is given using the largest singular value. The EW-TR approach, however, does not restrict the RIS channel to be a rank-one matrix and instead derives the solution using the trace operation. The SPGM approach adopts sum-path-gain maximization as the target of the RIS phase shift optimization \cite{Ning:2020}. The WMSE-up approach of \cite{Chen:2022} is a WMMSE-based technique, but the stationary point is derived for the upper bound of the WMSE. The IDM methodology minimizes the distance between the RIS-assisted channel and the optimal channel state \cite{Zhou:2021}. The array steering technique is a method that aligns the RIS phase shift vector with the directions of incoming and outgoing signals \cite{Hong:2022}.

\begin{table*}
    \centering
    \renewcommand{\arraystretch}{1.5}
    \begin{tabular}{|c|c|c|}
    \hline
         & \textbf{Algorithm}  & \textbf{Complexity} \\ \hline
        \multirow{2}{*}{\textbf{Beamformer}} & SVD + Water filling & $4MN^2+MLN+\frac{1}{2}N^2$ \\ \cline{2-3}
         & WMMSE & $M^3+N^3+2M^2N+6MN^2+MLN$ \\ \hline
        \multirow{6}{*} & GD & $2MN^2+M^2N+N^3+2L^2N^4+\frac{2}{3}LN^5+MLN^2$ \\ \cline{2-3}
        & EW-TR & $ML^2N+M^3L+4M^2LN$ \\ \cline{2-3}
        \textbf{RIS phase} & EW-SV & $ML^2N+2M^3L+4M^2LN$ \\ \cline{2-3}
        \textbf{optimization} & SPGM  & $M^2N+MLN^2+LN^3$ \\ \cline{2-3}
         & WMSE-up & $2MN+L$ \\ \cline{2-3}
         & IDM & $L^4$ \\ \hline
    \end{tabular}
    \caption{Computational complexity of RIS techniques for SU-MIMO systems.}
    \label{tab1}
\end{table*}

\begin{figure}[!t]
	\centering
	\includegraphics[width=0.95\columnwidth]{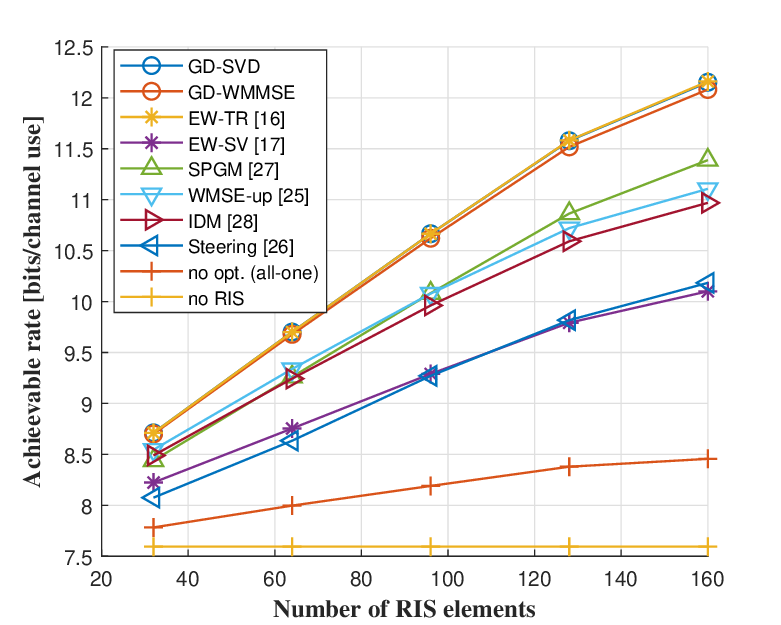}
	\caption{SU-MIMO achievable rates versus number of RIS elements, with 8 BS antennas and 4 UEs.}
	\label{fig:su_rate_vs_ris_elmts}
\end{figure}

\begin{figure}[!t]
	\centering
	\includegraphics[width=0.95\columnwidth]{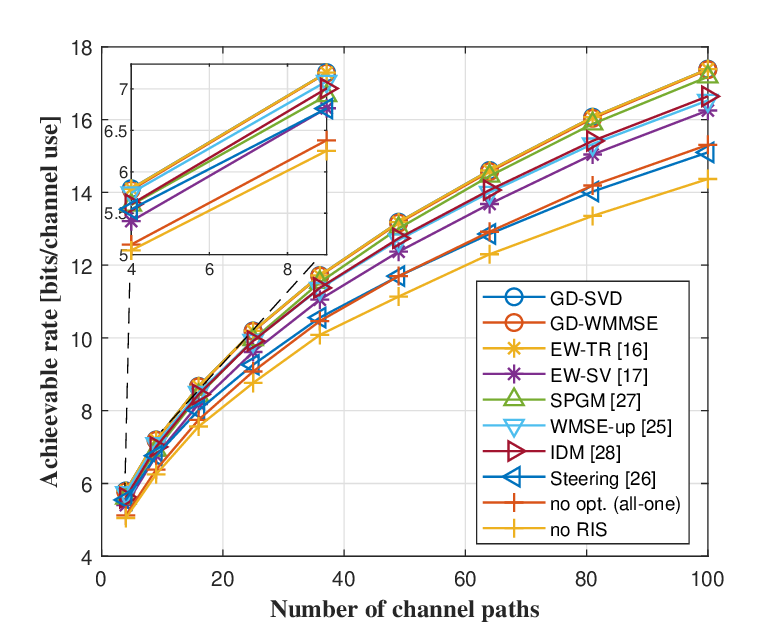}
	\caption{Achievable rates for SU-MIMO system techniques versus number of channel paths, 8 BS antennas, 4 UEs, 64 RIS elements.}
	\label{fig:su_rate_vs_k}
\end{figure}

In Fig.~\ref{fig:su_rate_vs_ris_elmts}, we evaluate the achievable rate for each technique as a function of the number of RIS elements. The proposed GD-SVD approach, which combines the GD method with the SVD beamformer, demonstrates the best performance together with EW-TR. There is a significant performance gap between EW-TR and EW-SV due to the rank-one channel model assumed by EW-SV. The SPGM approach uses the alternating direction method of multipliers (ADMM) for the suboptimal problem, and slowly approaches optimal performance as the number of RIS elements increases. Fig.~\ref{fig:su_rate_vs_ris_elmts} also contrasts the two WMMSE-based techniques: i) GD-WMMSE, which involves our proposed GD method based on the WMMSE beamformer, and ii) WMSE-up in \cite{Chen:2022}, based on a stationary point of the WMSE upper bound. GD-WMMSE significantly outperforms WMSE-up since the latter confines its search to the WMSE upper bound. IDM performs well with a small number of RIS elements, but its efficacy declines as the number of RIS elements increases since the lower bound used in IDM becomes less tight. The array steering technique is suitable for a large number of RIS elements, although it does not offer advantages when compared to the other techniques.

The superior performance of GD-SVD and EW-TR for different numbers of channel paths is clearly evident from Fig.~\ref{fig:su_rate_vs_k}. For sparse channels with a small number of paths, GD-WMMSE and WMSE-up also show performance close to GD-SVD. The lower bound derived in the IDM algorithm is tight for a small number of paths and it outperforms the SPGM-based algorithm. However, in rich scattering environments with many paths, SPGM achieves better performance. The array steering technique is also effective in sparse channel conditions, and while EW-SV exhibits consistent performance enhancement as the number of channel paths increases, it generally lags in performance compared to the other techniques.

\begin{figure}[!t]
	\centering
	\includegraphics[width=0.95\columnwidth]{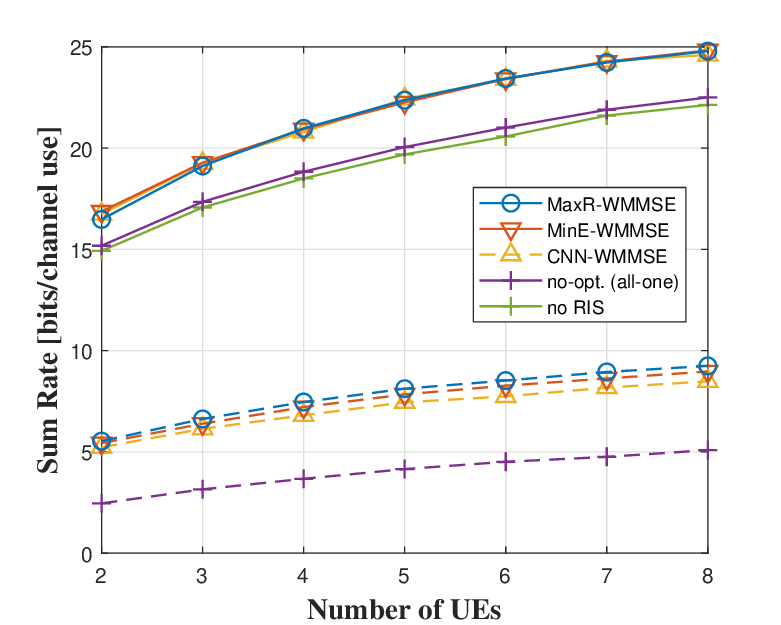}
	\caption{MU-MIMO sum-rate versus number of UEs, 8 BS antennas, 32 RIS elements, 4 UE antennas each.}
	\label{fig:rate_vs_ue}
\end{figure}

The computational complexity of the various techniques in the SU-MIMO case is detailed in Table~\ref{tab1}. In this table, the number of BS antennas is $M$, the number of UE antennas is $N$, and the number of RIS elements is $L$. For SU-MIMO systems, the SVD beamformer has lower complexity and achieves higher rate than the WMMSE beamformer as in Figs.~\ref{fig:su_rate_vs_ris_elmts}~and~\ref{fig:su_rate_vs_k}. For the RIS phase shift optimization in SU-MIMO systems, the GD method seems to have the highest complexity, proportional to $N^5$. However, IDM has highest order of complexity in $L$. Since the number of RIS elements is usually assumed to be much larger than the number of UE antennas, the complexity of the GD method may be tolerable for practical scenarios.

\subsection{Experiments for MU-MIMO Systems}
\label{sec5_2}
Since the SVD-based beamformer is not directly applicable for MU-MIMO systems, we focus on the RIS phase shift optimization algorithms based on the WMMSE beamformer: i) MaxR-WMMSE, maximizing the sum-rate, ii) MinE-WMMSE, minimizing the WMSE, and iii) CNN-WMMSE, learning the network to maximize the sum-rate. In Figs.~7--11 we investigate two scenarios, i.e., with and without the direct channel paths, where the solid lines are the experiments with the direct channel paths, and the dashed lines are the results without the direct channel paths.

In Fig.~\ref{fig:rate_vs_ue}, the sum-rate performance is analyzed in terms of the number of UEs. If direct channel paths exist, the RIS without optimization does not provide much gain, only 0.3326 bits per channel use (bpcu) higher than the no-RIS case. However, the improvement is about 2.44 bpcu for the WMMSE-based methods, a 7x improvement. The performance difference among the WMMSE-based methods is clear in the scenario without direct channel paths. For this scenario, MaxR-WMMSE provides the highest sum-rate although the gap is not significant.

\begin{figure}[!t]
	\centering
	\includegraphics[width=0.95\columnwidth]{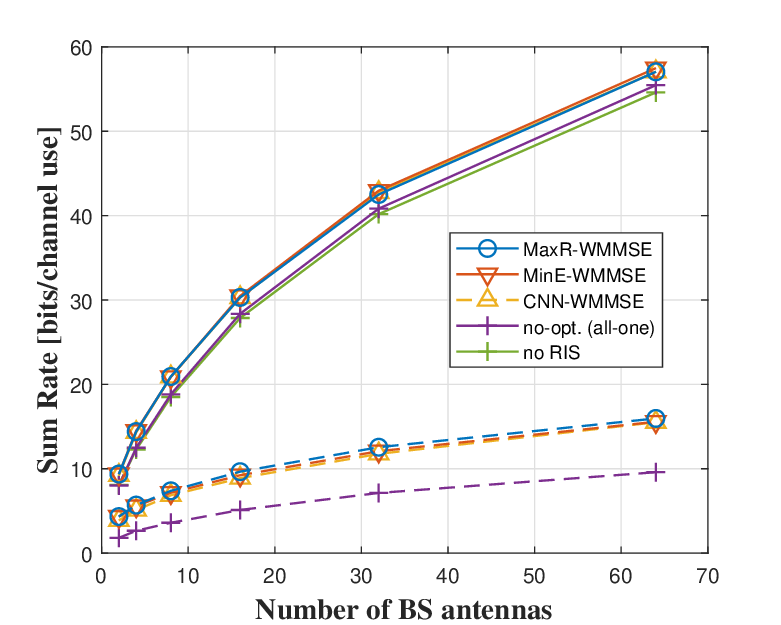}
	\caption{MU-MIMO sum-rate versus number of BS antennas, 4 UEs, 32 RIS elements, 4 UE antennas each.}
	\label{fig:rate_vs_bs_ants}
\end{figure}
\begin{figure}[!t]
	\centering
	\includegraphics[width=0.95\columnwidth]{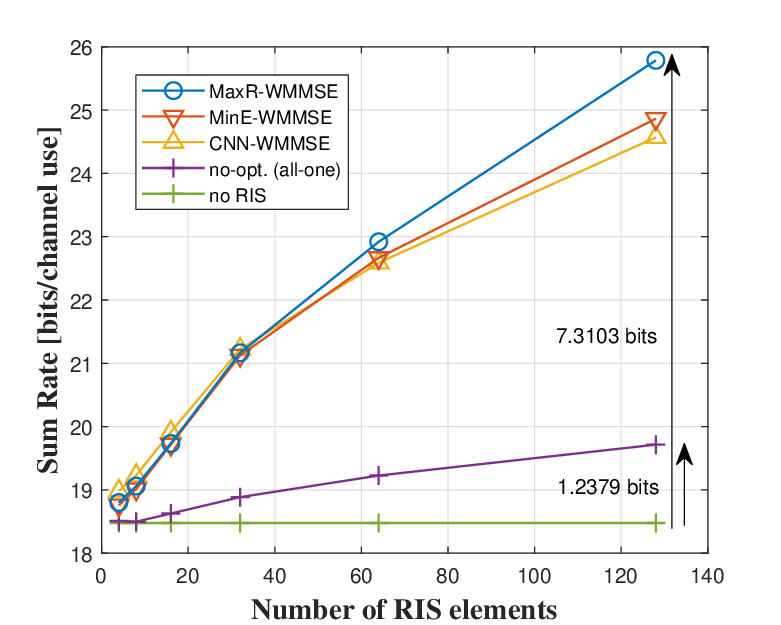}
	\caption{MU-MIMO sum-rate versus number of RIS elements, 8 BS antennas, 4 UEs with 4 antennas each.}
	\label{fig:rate_vs_ris_elmts}
\end{figure}
\begin{figure}[!t]
	\centering
	\includegraphics[width=0.95\columnwidth]{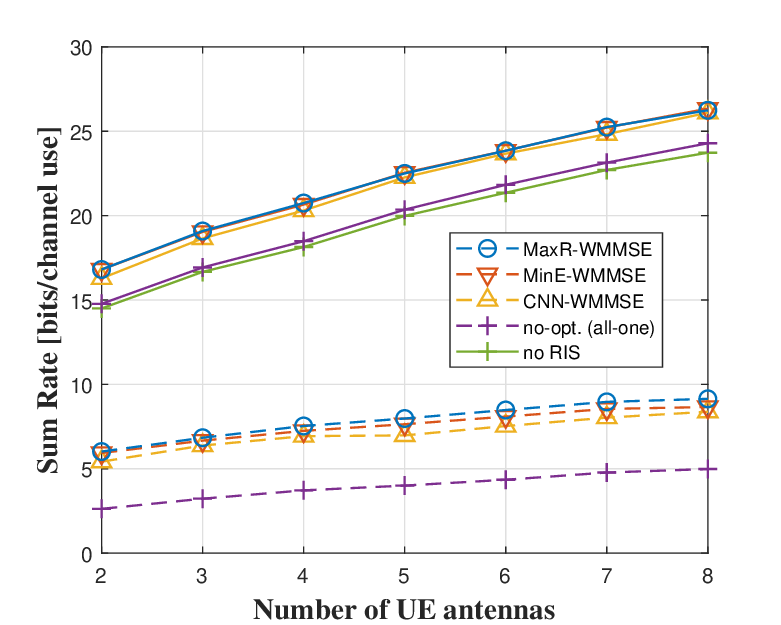}
	\caption{MU-MIMO sum-rate versus number of UE antennas, 8 BS antennas, 4 UEs, 32 RIS elements.}
	\label{fig:rate_vs_ue_ants}
\end{figure}
\begin{figure}[!t]
	\centering
	\includegraphics[width=0.95\columnwidth]{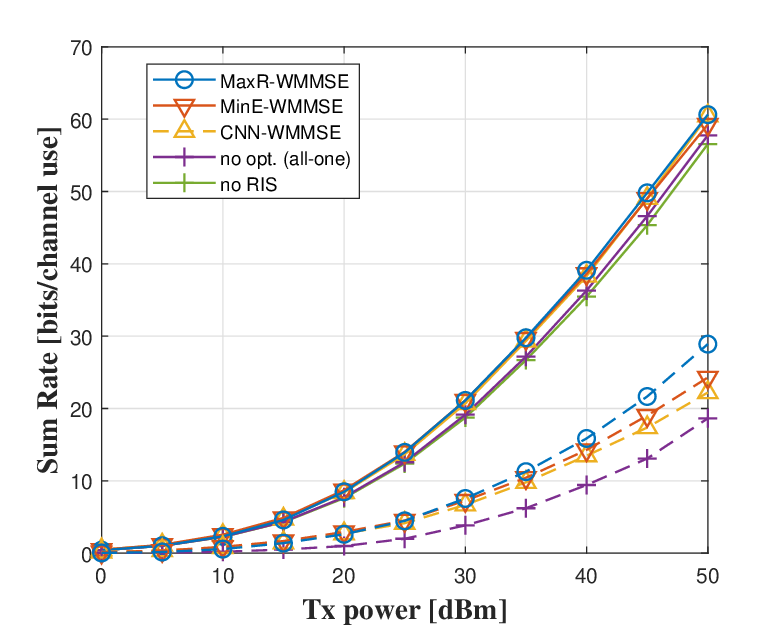}
	\caption{MU-MIMO sum-rates versus Tx power, 8 BS antennas, 4 UEs with 4 antennas each, 32 RIS elements.}
	\label{fig:rate_vs_tx_pw}
\end{figure}

\begin{table*}
    \centering
    \renewcommand{\arraystretch}{1.5}
    \begin{tabular}{|c|c|c|}
    \hline
         & \textbf{Algorithm}  & \textbf{Complexity} \\ \hline
         \textbf{Beamformer} & WMMSE & $K(M^3+KN^3+2KM^2N+6KMN^2)$ \\ \hline
        \multirow{2}{*} \textbf{\textbf{RIS phase}} & MaxR & $K(2KMN^2+KM^2N+L^3N^3+2L^2N^4+\frac{2}{3}LN^5+MLN^2)$ \\ \cline{2-3}
        \textbf{optimization}& MinE & $3K^2LN^3+2K^2MLN^2+L^3$ \\ \hline
    \end{tabular}
    \caption{Computational complexity of RIS techniques for MU-MIMO systems.}
    \label{tab2}
\end{table*}

Fig.~\ref{fig:rate_vs_bs_ants} depicts the sum-rate for various algorithms versus the number of BS antennas. All WMMSE-based approaches perform similarly for this case, and we see the same performance enhancement as the number of BS antennas increases. On the contrary, there is a significant difference in performance as the number of RIS elements varies, as shown in Fig.~\ref{fig:rate_vs_ris_elmts}. In particular, the performance of MaxR-WMMSE improves rapidly with a larger number of RIS elements. Unlike the case with an increasing number of UEs, we empirically observe that increasing the size of the RIS does not increase the number of local optima, and MaxR-WMMSE can effectively identify the correct gradient for the optimization by leveraging the derivative of the sum-rate when the number of RIS elements is large. Despite its simple structure in Fig.~\ref{fig:CNNstructure}, CNN-WMMSE delivers RIS gain comparable to MinE-WMMSE. However, as explained in Section~\ref{sec3_3}, CNN-WMMSE is associated with a significant overhead. There is a slight improvement in sum rate without optimization, but the gain is much lower than what can be obtained with the proposed algorithms.

We also compare the performance with the number of UE antennas and Tx power level in Figs.~\ref{fig:rate_vs_ue_ants}~and~\ref{fig:rate_vs_tx_pw}, respectively. The overall tendency for these cases is the same as in Fig.~\ref{fig:rate_vs_ue}, i.e., the performance of the WMMSE-based approaches is comparable when the direct channels exist, while MaxR-WMMSE exhibits the best performance without the direct channels. All these results highlight the superiority of MaxR-WMMSE in scenarios involving blockages.

\begin{figure}[!t]
    \centering
    \includegraphics[width=\columnwidth]{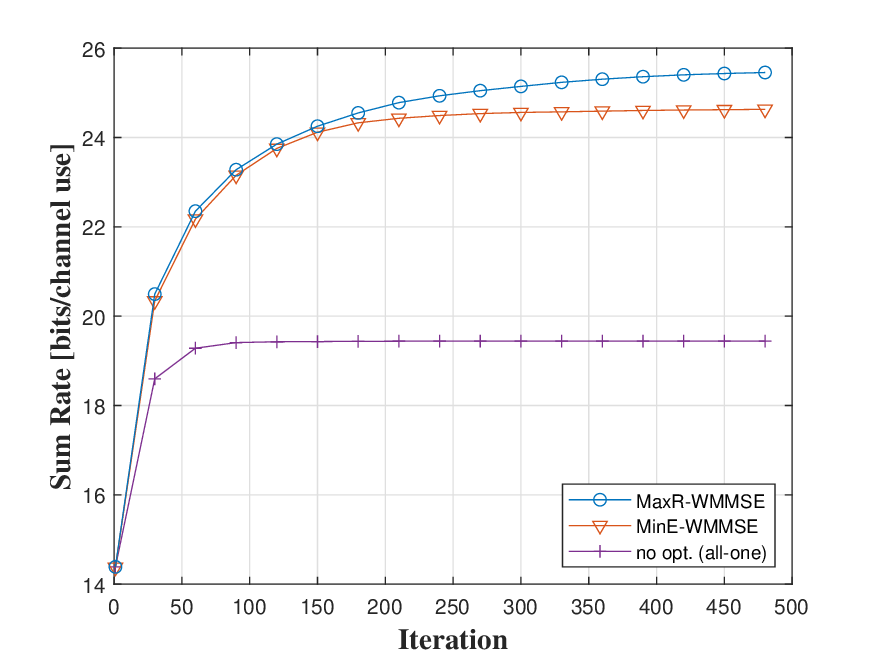}
    \caption{Convergence of RIS techniques for MU-MIMO systems, 8 antenna BS, 128 RIS elements, and 4 UEs with 4 antennas each.}
    \label{fig:convergence}
\end{figure}

Fig.~\ref{fig:convergence} illustrates algorithm convergence for MU-MIMO systems. As shown in Section~\ref{sec3_2}, MaxR-WMMSE converges and gives the highest sum-rate. Even though MinE-WMMSE does not guarantee a rate increase as in (40), the numerical result in Fig.~\ref{fig:convergence} nonetheless demonstrates that it converges. Note that the performance without RIS optimization also increases with each iteration since the WMMSE beamformer is iteratively updated.

Finally, we analyze the complexity of the considered MU-MIMO techniques in Table~\ref{tab2}. All the variables are the same as in Table I, while $K$ is the number of users. As shown in the table, MaxR-WMMSE requires higher computational complexity than MinE-WMMSE for MU-MIMO systems, and its complexity increases dramatically when the number of UE antennas increases. MaxR-WMMSE has a more reasonable complexity with a large number of RIS elements and small $N$.

\section{Conclusions}
\label{sec6}
In this paper, we first derived a tensor-based RIS system model that incorporates the RIS physical response. We then proposed a novel RIS phase shift optimization technique that maximizes the sum-rate using gradient descent. Numerical results showed that the proposed technique outperforms existing benchmarks in both SU-MIMO and MU-MIMO scenarios.

The tensor-based RIS system model is very general and it would be useful to develop more practical RIS-related techniques for these general scenarios. Actual RIS behavior will inevitably lead to RIS channels with rank higher than one, and thus the tensor-based RIS model is essential to achieving the best performance in practice.


{\appendices
    \section*{Appendix A}
	\section*{Proof of Lemma 1}
	Let the vector $\ba_n$ represent the $n$-th column of matrix $\bA$. For matrices $\bA$ and $\bB$ with proper dimensions, we have
    \begin{align}\label{eq:lemma_eq1}
        &\det(\bI+\bA^\mathrm{T}\bB\bA) \nonumber \\
        &=\det\begin{bmatrix}
            1+\ba_1^\mathrm{T}\bB\ba_1 & \ba_1^\mathrm{T}\bB\ba_2 & \cdots & \ba_1^\mathrm{T}\bB\ba_N \\
            \ba_2^\mathrm{T}\bB\ba_1 & 1+\ba_2^\mathrm{T}\bB\ba_2 & \cdots & \ba_2^\mathrm{T}\bB\ba_N \\
            \vdots & \vdots & \ddots & \vdots \\
            \ba_N^\mathrm{T}\bB\ba_1 & \ba_N^\mathrm{T}\bB\ba_2 & \cdots & 1+\ba_N^\mathrm{T}\bB\ba_N
        \end{bmatrix}, \nonumber \\
        &\stackrel{(a)}{=}(1+\ba_1^\mathrm{T}\bB\ba_1) \nonumber \\
        &\quad\times\det
        \left\{
        \begin{bmatrix}
            1+\ba_2^\mathrm{T}\bB\ba_2 & \ba_2^\mathrm{T}\bB\ba_3 & \cdots & \ba_2^\mathrm{T}\bB\ba_N \\
            \ba_3^\mathrm{T}\bB\ba_2 & 1+\ba_3^\mathrm{T}\bB\ba_3 & \cdots & \ba_3^\mathrm{T}\bB\ba_N \\
            \vdots & \vdots & \ddots & \vdots \\
            \ba_N^\mathrm{T}\bB\ba_2 & \ba_N^\mathrm{T}\bB\ba_3 & \cdots & 1+\ba_N^\mathrm{T}\bB\ba_N
        \end{bmatrix}\right. \nonumber \\
        &-\frac{1}{1+\ba_1^\mathrm{T}\bB\ba_1}\left.
        \begin{bmatrix}
            \ba_2^\mathrm{T}\bB\ba_1 \\ \ba_3^\mathrm{T}\bB\ba_1 \\ \vdots \\ \ba_N^\mathrm{T}\bB\ba_1
        \end{bmatrix}
        \begin{bmatrix}
            \ba_1^\mathrm{T}\bB\ba_2 & \ba_1^\mathrm{T}\bB\ba_3 & \cdots & \ba_1^\mathrm{T}\bB\ba_N^\mathrm{T}
        \end{bmatrix}
        \right\} \nonumber \\
        &=(1+\ba_1^\mathrm{T}\bP_1^\bot\ba_1) \nonumber \\
        &\quad\times\det
        \begin{bmatrix}
            1+\ba_2^\mathrm{T}\bP_2^\bot\ba_2 &
            \ba_2^\mathrm{T}\bP_2^\bot\ba_3 &
            \cdots &
            \ba_2^\mathrm{T}\bP_2^\bot\ba_N \\
            \ba_3^\mathrm{T}\bP_2^\bot\ba_2 &
            1+\ba_3^\mathrm{T}\bP_2^\bot\ba_3 &
            \cdots &
            \ba_3^\mathrm{T}\bP_2^\bot\ba_N \\
            \vdots & \vdots & \ddots & \vdots \\
            \ba_N^\mathrm{T}\bP_2^\bot\ba_2 &
            \ba_N^\mathrm{T}\bP_2^\bot\ba_3 &
            \cdots &
            1+\ba_N^\mathrm{T}\bP_2^\bot\ba_N 
        \end{bmatrix}
    \end{align}
    where $\bP_1^\bot\triangleq\bB$ and $\bP_2^\bot\triangleq\left(\bP_1^\bot-\frac{\bP_1^\bot\ba_1\ba_1^\mathrm{T}\bP_1^\bot}{\ba_1^\mathrm{T}\bP_1^\bot\ba_1}\right)$. The determinant lemma for block matrices leads to (a) in \eqref{eq:lemma_eq1}, which can be repeated as
    \begin{align}
        &\det(\bI+\bA^\mathrm{T}\bB\bA) \nonumber \\
        &=(1+\ba_1^\mathrm{T}\bP_1^\bot\ba_1)(1+\ba_2^\mathrm{T}\bP_2^\bot\ba_2) \nonumber \\
        &\ \ \times \det\begin{bmatrix}
            1+\ba_2^\mathrm{T}\bP_3^\bot\ba_2 &
            \ba_2^\mathrm{T}\bP_3^\bot\ba_3 &
            \cdots &
            \ba_2^\mathrm{T}\bP_3^\bot\ba_N \\
            \ba_3^\mathrm{T}\bP_3^\bot\ba_2 &
            1+\ba_3^\mathrm{T}\bP_3^\bot\ba_3 &
            \cdots &
            \ba_3^\mathrm{T}\bP_3^\bot\ba_N \\
            \vdots & \vdots & \ddots & \vdots \\
            \ba_N^\mathrm{T}\bP_3^\bot\ba_2 &
            \ba_N^\mathrm{T}\bP_3^\bot\ba_3 &
            \cdots &
            1+\ba_N^\mathrm{T}\bP_3^\bot\ba_N
        \end{bmatrix},
    \end{align}
    with $\bP_{m+1}^\bot=\bP_m^\bot-\frac{\bP_m^\bot\ba_m\ba_m^\mathrm{T}\bP_m^\bot}{1+\ba_m^\mathrm{T}\bP_m^\bot\ba_m}$. By completing the recursive extraction, the result in Lemma 1 is obtained as
    \begin{align}
        \det(\bI+\bA^\mathrm{T}\bB\bA)=\prod_{m=1}^M(1+\ba_m^\mathrm{T}\bP_m^\bot\ba_m),
    \end{align}
    which is a multiplication of scalar values.

    \section*{Appendix B}
    \section*{Proof of Theorem 1}
    A set of projections on the manifold along the gradient $(\nabla_{\bphi}\cR)^\ast$ can be expressed as
    \begin{align}
        \cC_{\bphi,(\nabla_{\bphi}\cR)^\ast}= \{\cP(\bphi+\beta(\nabla_{\bphi}\cR)^\ast)|\beta\in\mathbb{R}\},
    \end{align}
    where we define two distinct projections $\bphi^+,\bphi^-\in\cC_{\bphi,(\nabla_{\bphi}\cR)^\ast}$ as
    \begin{align}
        \begin{cases}
            \bphi^+= \cP(\bphi+\beta^+(\nabla_{\bphi}\cR)^\ast),\quad \beta^+<0,  \\
            \bphi^-= \cP(\bphi+\beta^-(\nabla_{\bphi}\cR)^\ast),\quad \beta^->0.
        \end{cases}
    \end{align}
    They allow the following representation:
    \begin{align}
        \begin{cases}
            \bphi^+= \bphi+\beta(\nabla_{\bphi}\cR)^\ast+\br^+,\\
            \bphi^-= \bphi+\beta(\nabla_{\bphi}\cR)^\ast+\br^-,
        \end{cases}
        \quad \beta > 0,
    \end{align}
    with
    \begin{align}
        \begin{cases}\br^+=\cP(\bphi+\beta^+(\nabla_{\bphi}\cR)^\ast)-(\bphi
        +\beta(\nabla_{\bphi}\cR)^\ast), \\
        \br^-=\cP(\bphi+\beta^-(\nabla_{\bphi}\cR)^\ast)-(\bphi+\beta(\nabla_{\bphi}\cR)^\ast).\end{cases} 
    \end{align}
    For any $\epsilon$ and $r_0=\lVert\beta(\nabla_{\bphi}\cR)^\ast\rVert$, we can find $\beta^+$ and $\beta^-$ satisfying $\lVert\br^+\rVert=r_0 + \epsilon$ and $\lVert\br^-\rVert=r_0 - \epsilon$. Then, it can be proved that, by the gradient property, there exists $\beta$ and $\epsilon$ for which $\cR(\bphi^-)\geq \cR(\bphi)\geq\cR(\bphi^+)$ holds. For such $\beta$ and $\epsilon$, we can always find $\beta^\star$ satisfying $\bphi^-=\cP(\bphi+\beta^\star(\nabla_{\bphi}\cR)^\ast)$. Therefore, gradient descent on the manifold always increases the rate with proper $\beta^\star$.

	\section*{Appendix C}
    \section*{Derivation of MSE}
    By the definition, the MSE can be written as
    \begin{align}\label{eq:WMSE1}
    	\mathbb{E}[\bE_k]&=\mathbb{E}\Bigg[\left\{\bA_k\left(\bH_k^\mathrm{H}\sum_{i=1}^K\bB_i\bd_i+\bn_k\right)-\bd_k\right\} \nonumber \\
    	&\qquad\qquad\times\left\{\bA_k\left(\bH_k^\mathrm{H}\sum_{i=1}^K\bB_i\bd_i+\bn_k\right)-\bd_k\right\}^\mathrm{H}\Bigg] \nonumber \\
    	&\stackrel{(a)}{=}\sum_{i=1}^K\bA_k\bH_k^\mathrm{H}\bB_i\bB_i^\mathrm{H}\bH_k\bA_k^\mathrm{H}-\bA_k\bH_k^\mathrm{H}\bB_k \nonumber \\
    	&\qquad\qquad-\bB_k^\mathrm{H}\bH_k\bA_k^\mathrm{H}+\bI_N+\sigma_n^2\bA_k\bA_k^\mathrm{H},
    \end{align}
    where (a) holds because all symbols are assumed to be independent with zero mean and unit variance, which gives $\mathbb{E}[\bd\bd^\mathrm{H}]=\bI_{KN}$. With the definition of $\bH_k$ in \eqref{eq:RISchannel}, the MSE in \eqref{eq:WMSE1} can be further expanded as
    \begin{align}\label{eq:WMSE2}
    	\mathbb{E}[\bE_k]&=\sum_{i=1}^K\bA_k(\bH_{\mathrm{d},k}+[\![\boldsymbol{\cH}_{\mathrm{R},k}\times_2\bphi]\!])^\mathrm{H}\bB_i \nonumber \\
    	&\quad\qquad\times\bB_i^\mathrm{H}(\bH_{\mathrm{d},k}+[\![\boldsymbol{\cH}_{\mathrm{R},k}\times_2\bphi]\!])\bA_k^\mathrm{H} \nonumber \\
    	&\quad-\bA_k(\bH_{\mathrm{d},k}+[\![\boldsymbol{\cH}_{\mathrm{R},k}\times_2\bphi]\!])^\mathrm{H}\bB_k \nonumber \\
    	&\quad-\bB_k^\mathrm{H}(\bH_{\mathrm{d},k}+[\![\boldsymbol{\cH}_{\mathrm{R},k}\times_2\bphi]\!])\bA_k^\mathrm{H}+\bI_N+\sigma_n^2\bA_k\bA_k^\mathrm{H} \nonumber \\
    	&=\sum_{i=1}^K\bA_k[\![\boldsymbol{\cH}_{\mathrm{R},k}\times_2\bphi]\!]^\mathrm{H}\bB_i\bB_i^\mathrm{H}[\![\boldsymbol{\cH}_{\mathrm{R},k}\times_2\bphi]\!]\bA_k^\mathrm{H} \nonumber \\
    	&\quad+\sum_{i=1}^K\bA_k[\![\boldsymbol{\cH}_{\mathrm{R},k}\times_2\bphi]\!]^\mathrm{H}\bB_i\bB_i^\mathrm{H}\bH_{\mathrm{d},k}\bA_k^\mathrm{H} \nonumber \\
    	&\quad+\sum_{i=1}^K\bA_k\bH_{\mathrm{d},k}^\mathrm{H}\bB_i\bB_i^\mathrm{H}[\![\boldsymbol{\cH}_{\mathrm{R},k}\times_2\bphi]\!]\bA_k^\mathrm{H} \nonumber \\
    	&\quad-\bA_k[\![\boldsymbol{\cH}_{\mathrm{R},k}\times_2\bphi]\!]^\mathrm{H}\bB_k-\bB_k^\mathrm{H}[\![\boldsymbol{\cH}_{\mathrm{R},k}\times_2\bphi]\!]\bA_k^\mathrm{H}+\cO_\bphi,
    \end{align}
    where $\cO_\bphi$ is an aggregated term that does not depend on $\bphi$. Defining the effective channels $\tilde{\boldsymbol{\cH}}_{k,i}$ and $\tilde{\bH}_{k,i}$ with the combiner $\bA_k$ for the $k$-th UE and the beamformer $\bB_i$ for the $i$-th UE as
    \begin{align}
    \tilde{\boldsymbol{\cH}}_{k,i}&\triangleq(\boldsymbol{\cH}_{\mathrm{R},k}\times_1\bB_i^\ast)\times_3\bA_k^\mathrm{H} \\\tilde{\bH}_{k,i}&\triangleq\bB_i^\mathrm{H}\bH_{\mathrm{d},k}\bA_k^\mathrm{H},
    \end{align}
    the MSE for the $k$-th UE can be expressed as
    \begin{align}\label{eq:WMMSE3}
    	&\mathbb{E}[\bE_k]=\sum_{i=1}^K[\![\tilde{\boldsymbol{\cH}}_{k,i}\times_2\bphi]\!]^\mathrm{H}[\![\tilde{\boldsymbol{\cH}}_{k,i}\times_2\bphi]\!] \nonumber \\
    	&\qquad\quad +\sum_{i=1}^K[\![\tilde{\boldsymbol{\cH}}_{k,i}\times_2\bphi]\!]^\mathrm{H}\tilde{\bH}_{k,i}+\sum_{i=1}^K\tilde{\bH}_{k,i}^\mathrm{H}[\![\tilde{\boldsymbol{\cH}}_{k,i}\times_2\bphi]\!] \nonumber\\
    	&\qquad\quad-[\![\tilde{\boldsymbol{\cH}}_{k,k}\times_2\bphi]\!]^\mathrm{H}-[\![\tilde{\boldsymbol{\cH}}_{k,k}\times_2\bphi]\!]+\cO_\bphi.
    \end{align}
    
    \noindent Using Property 1 in Section~\ref{sec3_2}, the MSE in \eqref{eq:WMMSE3} can be further rewritten as
    \begin{align}
    	&\mathbb{E}[\bE_k]=\sum_{i=1}^K\begin{bmatrix}\bar{\bR}_{k,i,1}^\mathrm{H} & \bar{\bR}_{k,i,2}^\mathrm{H} & \cdots \end{bmatrix}(\bI\otimes\bphi^\ast) \nonumber \\
    	&\qquad\qquad\quad\times(\bI\otimes\bphi^\mathrm{T})\begin{bmatrix}\bar{\bR}_{k,i,1}^\mathrm{T} & \bar{\bR}_{k,i,2}^\mathrm{T} & \cdots\end{bmatrix}^\mathrm{T} \nonumber \\
    	&\qquad\quad+\sum_{i=1}^K(\bI\otimes\bphi^\mathrm{H})\begin{bmatrix}\tilde{\bR}_{k,i,1} & \tilde{\bR}_{k,i,2} & \cdots\end{bmatrix}^\mathrm{H}\tilde{\bH}_{k,i} \nonumber \\
    	&\qquad\quad+\sum_{i=1}^K\tilde{\bH}_{k,i}^\mathrm{H}\begin{bmatrix}\tilde{\bR}_{k,i,1} & \tilde{\bR}_{k,i,2} & \cdots \end{bmatrix}(\bI\otimes\bphi) \nonumber \\
    	&\qquad\quad-(\bI\otimes\bphi^\mathrm{H})\begin{bmatrix}\tilde{\bR}_{k,k,1} & \tilde{\bR}_{k,k,2} & \cdots\end{bmatrix}^\mathrm{H} \nonumber \\
    	&\qquad\quad-\begin{bmatrix}\tilde{\bR}_{k,k,1} & \tilde{\bR}_{k,k,2} & \cdots\end{bmatrix}(\bI\otimes\bphi)+\cO_\bphi,
    \end{align}
    where $\bar{\bR}_{k,i,n}=[\![\tilde{\boldsymbol{\cH}}_{k,i}(n,:,:)]\!]$ and $\tilde{\bR}_{k,i,n}=[\![\tilde{\boldsymbol{\cH}}_{k,i}(:,:,n)]\!]$ are shrunk matrices.    
}

\vfill
	
\end{document}